\documentclass[sigplan]{acmart}

\acmConference[PL'18]{ACM SIGPLAN Conference on Programming Languages}{January 01--03, 2018}{New York, NY, USA}
\acmYear{2018}
\acmISBN{} 
\acmDOI{} 
\startPage{1}

\setcopyright{none}

\usepackage{booktabs}   
\usepackage{subcaption} 

\usepackage{verbatim}
\usepackage{algorithm}
\usepackage{algpseudocode}
\usepackage{mathtools}
\usepackage{centernot}
\usepackage{tikz}
\usepackage{xspace}
\usepackage{xcolor}
\usepackage{multirow}

\newcommand{\GG}{\mathcal{G}}
\newcommand{\LL}{\mathcal{L}}
\newcommand{\D}{\mathcal{D}}

\newcommand{\T}{\mathcal{T}}

\newcommand{\OO}{\mathcal{O}}
\newcommand{\Q}{\mathcal{Q}}
\newcommand{\N}{\mathcal{N}}
\newcommand{\PC}{\mathcal{P}}
\newcommand{\PP}{\mathcal{P}}

\newcommand{\bool}{\mathbb{B}}

\newcommand{\semantics}[1]{{\llbracket#1\rrbracket}}
\newcommand{\holes}[1]{\textsf{holes}(#1)}
\newcommand{\nodes}[1]{\textsf{nodes}(#1)}

\newcommand{\leaves}[1]{\textsf{leaves}(#1)}
\newcommand{\internal}[1]{\textsf{internal}(#1)}
\newcommand{\true}{\textsf{true}}
\newcommand{\false}{\textsf{false}}
\newcommand{\score}[1]{\textsf{score}(#1)}
\newcommand{\scorehat}[1]{\widehat{\textsf{score}}(#1)}
\newcommand{\scoretilde}[1]{\widetilde{\textsf{score}}(#1)}
\newcommand{\complete}[1]{\overline{#1}}

\newcommand{\toolname}{\textsc{iSQL}\xspace}

\begin{document}

\title{Synthesizing Queries via Interactive Sketching}         


\author{Osbert Bastani}
\affiliation{\institution{University of Pennsylvania}
    \country{USA}}
\email{obastani@seas.upenn.edu}

\author{Xin Zhang}
\affiliation{\institution{Peking University}
    \country{China}}
\email{xin@pku.edu.cn}

\author{Armando Solar-Lezama}
\affiliation{\institution{MIT}
    \country{USA}}
\email{asolar@csail.mit.edu}

\begin{abstract}
We propose a novel approach to program synthesis, focusing on synthesizing database queries. At a high level, our proposed algorithm takes as input a sketch with soft constraints encoding user intent, and then iteratively interacts with the user to refine the sketch. At each step, our algorithm proposes a candidate refinement of the sketch, which the user can either accept or reject. By leveraging this rich form of user feedback, our algorithm is able to both resolve ambiguity in user intent and improve scalability. In particular, assuming the user provides accurate inputs and responses, then our algorithm is guaranteed to converge to the true program (i.e., one that the user approves) in polynomial time. We perform a qualitative evaluation of our algorithm, showing how it can be used to synthesize a variety of queries on a database of academic publications.

\end{abstract}




\maketitle

\section{Introduction}
\label{sec:intro}

Program synthesis has emerged as a promising way to help users write programs---e.g., it has been leveraged to generate highly optimized bit manipulation programs~\cite{solar2005programming,solar2006combinatorial}, string processing programs~\cite{gulwani2011automating,polozov2015flashmeta}, and database queries~\cite{li2014constructing,yaghmazadeh2017sqlizer,feng2017component,wang2017synthesizing}. These techniques enable the user to focus on specifying their high-level intent. For example, one approach is for the user to provide a logical specification such as a sketch annotated with logical constraints~\cite{solar2005programming,solar2006combinatorial}. Then, the synthesizer automatically searches the space of programs to try and identify one that satisfies the given specification.

A key challenge is that the user may be unable to provide a precise, logical specification. One reason is that they may be an end-user who knows exactly what task they want to achieve, but is unfamiliar with logical specifications. In such cases, existing systems allow the user to provide \emph{ambiguous specifications} such as input-output examples~\cite{gulwani2011automating,polozov2015flashmeta,feser2015synthesizing,feng2017component,wang2017synthesizing} or natural language descriptions~\cite{li2014constructing,yaghmazadeh2017sqlizer}.

An alternative reason is that the user is familiar with logical specification, but uncertain about the task they want to achieve. For example, suppose a user wants to write a query to extract author names from a large database of academic publications. They may be uncertain about the task at hand---e.g., they may be unfamiliar with the database schema, so they are unable to write down a precise logical specification or even an ambiguous one. Alternatively, the task may be inherently uncertain---e.g., the user might want to run the query on a database whose schema changes over time, or on multiple databases with different schemas. In these cases, the user may want to provide an \emph{uncertain specification}---i.e., a precise, logical specification of not only their knowledge of their task, but also their uncertainty about their task.

We propose a specification language and corresponding program synthesis algorithm that enables users to express uncertain specifications. We assume the user is knowledgeable and can write both programs, but doing so requires significant effort to first resolve their uncertainty about their task. We focus on the setting of a user who is a data scientist trying to write a query to extract data from a large database. Their uncertainty may be because they are unfamiliar with the database schema, or because they want to run the same query on multiple databases with different schemas.

To use our system, the user provides a specification in the form of a sketch, which includes the structure of the query (including all select and project operations), but can have holes corresponding to tables (to be filled using a table constant or a sequence of inner-join operations) and columns (to be filled with a column name). In other words, the key portion of the query that the user can leave unspecified is the sequence of tables should be joined together to construct the desired flat table. The search space of such sequences can be very large---many datasets have dozens of tables and it is exponential in the number of tables to be joined.

To express their uncertainty, our language supports specifications in the form of soft constraints. In particular, the user can include soft constraints on expressions in the sketch; these constraints should encode the user's expectations about the value that should be obtained by evaluating the expression once all holes in the sketch have been filled. For example, to specify that a column should contain author names, the user might include a soft constraint saying that the values in that column are strings, some of which contain ``Church''.

Then, our synthesis algorithm fills the holes in the provided sketch, relying critically on user interaction to resolve uncertainty. We assume the user can recognize the correct query when they see it. This assumption may seem at odds with the fact that the user is uncertain about their intent, but manually resolving ambiguity can require significantly more effort than being actively guided through the process. In particular, our algorithm actively focuses the user's efforts on understanding portions of the database that are relevant to resolving the uncertainty they specified in their sketch.

\begin{figure}
\small
\begin{tabular}{rl}
\toprule
\multicolumn{1}{c}{{\bf aid}} & \multicolumn{1}{c}{{\bf name}} \\
\midrule
0 & Alan M. Turing \\
1 & Alonzo Church \\
\bottomrule
\multicolumn{2}{c}{{\bf authors}}
\end{tabular}
\hspace{0.3in}
\begin{tabular}{rr}
\toprule
\multicolumn{1}{c}{{\bf aid}} & \multicolumn{1}{c}{{\bf pid}} \\
\midrule
0 & 0 \\
0 & 1 \\
1 & 2 \\
\bottomrule
\multicolumn{2}{c}{{\bf writes}}
\end{tabular}
\hspace{0.3in}
\begin{tabular}{rlr}
\toprule
\multicolumn{1}{c}{{\bf pid}} & \multicolumn{1}{c}{{\bf title}} & \multicolumn{1}{c}{{\bf year}} \\
\midrule
0 & Computability and $\lambda$-definability & 1937 \\
1 & Intelligent machinery & 1948 \\
2 & A set of postulates for the foundation of logic & 1932 \\
\bottomrule
\multicolumn{3}{c}{{\bf publications}}
\end{tabular}
\caption{Example of a database of computer science publications and their authors. This database includes three tables---one of authors (``authors''), one of published papers (``publications''), and one that links the previous two (``writes'').}
\label{fig:exdb}
\end{figure}

At a high level, our algorithm keeps track of a sketch (i.e., a program with holes), which is initialized to be the given sketch. Then, it iteratively interacts with the user to fill the holes in the sketch. At each iteration, our algorithm proposes a candidate \emph{refinement} of the current sketch (i.e., proposes to fill a single hole with some expression) to the user, who either accepts the refinement if it matches the true program or rejects it otherwise. The added expression may contain new holes that must be filled in subsequent iterations. If the proposed refinement is accepted, then the current sketch is updated; otherwise, the current sketch remains unchanged. Either way, our algorithm continues to ask additional questions until the sketch is concrete (i.e., there are no remaining holes), at which point it is returned by our algorithm. We assume the user always correctly accepts or rejects the candidate refinement; in practice, they can backtrack if they realize their choices are incorrect.

Our algorithm enjoys two key advantages that derive from the rich feedback that it solicits from the user. First, it is guaranteed to find the true program (assuming the structure of the sketch is correct and the user always answers accurately). In particular, the true program must be attainable from some sequence of refinements of the given sketch, and the user affirms each refinement made by the algorithm, including the final program. This guarantee holds even if the constraints provided by the user are underspecified---i.e., it implicitly resolves ambiguity in the specified user intent. Of course, better specifications can lead to faster convergence.

Second, our algorithm is guaranteed to identify the true program in a number of iterations that is polynomial in the size of the true program. First, each iteration of our algorithm is efficient since our algorithm only has to search over the set of possible refinements, which is polynomial in the size of the sketch. Furthermore, our algorithm is guaranteed to identify the true program in a polynomial number of iterations. This guarantee holds for any choice of questions; in practice, we try to minimize the overall number of questions asked.

The key challenge is designing a space of possible refinements such that the user can quickly and accurately respond to questions. Since we assume the user is a knowledgeable programmer, we assume they can recognize the correct program by inspecting both the program code and the data---e.g., in our setting, the user bases their decision on both the query and the information in the database. The key advantage is that they only need to understand the portions of the database that are relevant to questions asked by our algorithm. Assuming our algorithm asks good questions, then we can significantly reduce the user's workload.

To respond to a question, the user must be able to determine whether the proposed refinement is correct without seeing the whole program. Intuitively, the refinement should always be a concrete transformation of the input that the user can inspect and validate---e.g., a key refinement our algorithm can propose is to replace a hole with an inner join of a concrete table with another hole. Thus, the user only needs to check that the concrete table is part of the sequence of inner-join operations to construct the desired flat table.

Finally, we evaluate our approach on a dataset of queries to a database of academic publications~\cite{li2014constructing}. We perform a small-scale user study to determine whether users can write reasonable sketches, as well as interact with our synthesis algorithm to refine a sketch. We find that 80\% of sketches are successful, and 93\% of questions were answered correctly, suggesting that users can use our system even with very little experience. Furthermore, we show that for all of our sketches, our algorithm quickly converges to the true query.

In summary, our contributions are:
\begin{itemize}
\item A novel formulation of program synthesis that interacts with the user to iteratively refine a user-provided specification in the form of a sketch (Section~\ref{sec:prob}).
\item An algorithm based on this approach in the context of synthesizing database queries (Section~\ref{sec:alg}).
\item An implementation of our algorithm in a tool called \toolname (Section~\ref{sec:impl}),
\footnote{\toolname stands for ``interactive SQL''; we pronounce it ``icicle''.}
and an evaluation that illustrates how our approach can be used to synthesize a variety of database queries (Section~\ref{sec:eval}).
\end{itemize}

\begin{figure}
\centering\scriptsize
\begin{tabular}{l}
\toprule\\
$\begin{array}{l}
\verb+SELECT ??c_name:column+ \\
\verb+FROM (??t:table +{\color{red}\texttt{\{(contains ??c\_name:column ".*Church.*")}} \\
\verb+                 +{\color{red}\texttt{AND (1900 <= ??c\_year:column <= 2020)\}}}\verb+)+ \\
\verb+WHERE ??c_year:column = 1948+
\end{array}$ \\\\
\multicolumn{1}{c}{{\bf sketch}}
\\\midrule\\
$\begin{array}{l}
\verb+??t:table+
\end{array}
\xRightarrow{*}
\begin{array}{l}
\verb+authors+ \\
\verb+INNER-JOIN ??t_new:table+ \\
\verb+ON ??c_new0:column = ??c_new1:column+
\end{array}
$ \\\\
\multicolumn{1}{c}{{\bf candidate production sequence}}
\\\midrule\\
$\begin{array}{l}
\verb+SELECT ??c_name:column+ \\
\verb+FROM (authors+ \\
\verb+      INNER-JOIN ??t_new:table+ \\
\verb+      ON ??c_new0:column = ??c_new1:column+ \\
\verb+      +{\color{red}\texttt{\{(contains ??c\_name:column ".*Church.*")}} \\
\verb+       +{\color{red}\texttt{AND (1900 <= ??c\_year:column <= 2020)\}}}\verb+)+ \\
\verb+WHERE ??c_year:column = 1948+
\end{array}$ \\\\
\multicolumn{1}{c}{{\bf refinement}}
\\\midrule\\
$\begin{array}{l}
\verb+SELECT name+ \\
\verb+FROM (authors+ \\
\verb+      INNER-JOIN (writes+ \\
\verb+                  INNER-JOIN publications+ \\
\verb+                  ON writes.pid = publications.pid)+ \\
\verb+      ON authors.aid = writes.aid)+ \\
\verb+      +{\color{red}\texttt{\{(contains name ".*Church.*") AND (1900 <= year <= 2020)\}}}\verb+)+ \\
\verb+WHERE publications.year = 1948+
\end{array}$ \\\\
\multicolumn{1}{c}{{\bf completion}}
\\\bottomrule
\end{tabular}
\caption{Examples illustrating the concepts used in our algorithm. The goal is to select authors who published papers in 1948. The user provides the sketch (first row). Then, our algorithm proposes a candidate sequence of productions to use to refine the sketch (second row). If accepted by the user, the productions are applied to the sketch to obtain the refined sketch (third row). This interactive process continues iteratively until the sketch has no more holes (last row). Soft constraints are shown in red.}
\label{fig:ex}
\end{figure}

\section{Overview}
\label{sec:overview}

As a motivating example, consider a user (e.g., a data scientist) who wants to query a database of academic publications, including information such as conferences/journals, titles, abstracts, authors, citations, etc. Their goal may be to perform data exploration, to compute some basic statistics of the data, or to construct a flat table that will be used to train a machine learning model. Potential queries include which authors were active in a given year (e.g., published a paper in 1948), which authors have the most citations, which authors have cited a given academic, etc.

While the user may be knowledgeable of SQL, they may not be familiar with the schema of the database. For example, many databases are poorly documented, including ones used for data science. As a consequence, it is challenging for the user to write the desired database query from scratch, since they have to spend time reading the database documentation and understanding its schema. However, if they are shown a candidate query, then they can easily check the corresponding tables and columns in the database to determine whether it is correct. This problem arises in data science tasks, where the user is a data scientist who is exploring a number of large datasets with the goal of deriving some kind of insight from the data. Many of these datasets are unfamiliar to the data scientist and poorly documented, making it time consuming to identify which tables and columns to use to fill in different parts of the desired query.

As a concrete example, consider the database shown in Figure~\ref{fig:exdb}, which includes three tables---one containing authors, one containing computer science publications, and a third that links the previous two. Suppose the user wants to select all authors who published a paper in 1948. To do so, the user can use the query on the last line of Figure~\ref{fig:ex}. In this query, the \texttt{writes} table relates authors to publications; thus, the query joins \texttt{publications} with \texttt{writes}, and then joins the result with \texttt{authors} to obtain a flat table with both authors and their publications. Then, this query selects the authors that have a publication in the year 1948.

\subsection{Initial User Specification}

We assume that the user is familiar with the query language (e.g., SQL) and knows what the desired data should look like (e.g., publication years are mostly between 1900 and 2020) but is unfamiliar with the database schema and does not know where in the database the desired information is located. Furthermore, we assume the database is large and possibly poorly documented, which is typical of many datasets used in data science tasks. In particular, the user is capable of writing a skeleton of the query (e.g., the structure of the output they ultimately want to construct, and any aggregation operators they want to apply), but does not know which tables and columns to use inside this query.

In our example, we expect that the user knows the structure of the outermost select statement (which combines selecting rows with year 1948 and a project that retains only the author name column). The key challenge is that they do not know how to construct the flat table that includes both authors and publications. More precisely, the key challenge is that the user does not know the sequence of tables to inner-join to obtain this flat table. Our algorithm is designed to help the user discover this sequence.

\paragraph{Sketches.}

To this end, we assume the user is able to provide the sketch shown on the first row of Figure~\ref{fig:ex}. This sketch outlines the structure of the select and project operators to be applied to the flat table. The flat table, along with the columns to be selected and projected, are left as holes in this sketch. In particular, there are three holes in this sketch:
\begin{itemize}
\item The hole \texttt{??c\_name:column} is named \texttt{c\_name}, has type \texttt{column}, and corresponds to the unknown author name column.
\item The hole \texttt{??t:table} is named \texttt{t}, has type \texttt{table}, and corresponds to the unknown flat table.
\item The hole \texttt{??c\_year:column} is named \texttt{c\_year}, has type \texttt{column} and corresponds to the unknown publication year column.
\end{itemize}
In general, holes either have type \texttt{column} or \texttt{table}. The names associated with holes are used to link different holes in the sketch that are known to have the same value---e.g., in the example sketch, there are two holes named \texttt{c\_year}, which indicates that they must be filled using the same column.

\begin{figure}
\centering\small
\begin{tabular}{rlrlr}
\toprule
\multicolumn{1}{c}{{\bf aid}} & \multicolumn{1}{c}{{\bf name}} & \multicolumn{1}{c}{{\bf pid}} & \multicolumn{1}{c}{{\bf title}} & \multicolumn{1}{c}{{\bf year}} \\
\midrule
0 & Alan M. Turing & 0 & ... & 1937 \\
0 & Alan M. Turing & 1 & ... & 1948 \\
1 & Alonzo Church & 2 & ... & 1932 \\
\bottomrule
\end{tabular}
\caption{The table obtained by evaluating the expression \texttt{(authors INNER-JOIN ...)} in the complete program on the last line of Figure~\ref{fig:ex} on the database in Figure~\ref{fig:exdb}.}
\label{fig:exdbex}
\end{figure}

\paragraph{Soft constraints.}

In addition, the user can also provide specifications that encode how the sketch should be filled. Unlike traditional specifications, which provide hard constraints on the semantics of the sketch, these specifications are soft constraints that encode expectations that the user has about likely properties of the semantics. These soft constraints can be used to assign a score to a concrete program that indicates how well the program matches the user's expectation.
\footnote{We use soft constraints since the user might be uncertain about the actual contents of the database; we could easily include hard constraints if desired.}

In our example sketch, the user has provided three soft constraints on the expected semantics. In the context of database queries, these constraints encode the user's expectations about the properties of the data in a table constructed while evaluating the query. For example, consider the portion
\begin{align*}
&\texttt{??t\_new:table} \\
&\texttt{\{(contains ??c\_name:column ".*Church.*") ...\}}
\end{align*}
of the sketch. This portion consists of a hole \texttt{??t\_new:table} with name \texttt{t\_new} and type \texttt{table}. The soft constraint is the expression appearing in the curly braces \texttt{\{...\}}. This constraint applies to the preceding expression---in this case, the hole named \texttt{t\_new}. Semantically, it maps the table $t$ obtained by evaluating the preceding expression to a real-valued score $s\in\mathbb{R}$. In our example, the soft constraint $\textsf{contains}(c,r)$ maps the table $t$ to $1$ if $c$ contains a string matching the regular expression $r$, and $0$ otherwise. Intuitively, this constraint encodes the user expectation that the table contains a column of strings, and that some of these strings (which should be author names) contain the substring ``Church''. This constraint is soft because they are interpreted in a way that does not prune a possible program just because it is violated.

We evaluate the soft constraint in the context of a \emph{completion} of the sketch---i.e., a concrete program obtained by filling all the holes in the sketch with concrete expressions. Then, the expression filling the hole named \texttt{t\_new} evaluates to some table $t$; then, we apply the soft constraint to $t$ to obtain a score $s$. For example, in Figure~\ref{fig:ex}, the concrete program on the last line is a completion of the sketch on the first line. In this completion, the expression above becomes
\begin{align*}
&\texttt{(authors INNER-JOIN ...)} \\
&\texttt{\{(contains name ".*Church.*") ...\}}
\end{align*}
in which case \texttt{(contains name ".*Church.*")} evaluates to $1$. In particular, the inner-join evaluates to the table shown in Figure~\ref{fig:exdbex}. Then, the soft constraint says the user expects that some of the values in the \texttt{name} column of this table contain the substring ``Church''. Since one of the values in this column satisfies this property, we assign score $1$.

Taken together, we can assign a score to a completion of a sketch by evaluating the completion, evaluating the soft constraints, and summing the resulting scores. Higher scoring completions correspond to concrete programs that are more likely according to the user-provided specification.

\subsection{Interactive Program Synthesis Algorithm}

Our algorithm interacts with the user to determine how to fill the holes in the given sketch. It keeps track of a sketch $P$, which is initialized to the given sketch. At each iteration, it proposes a candidate \emph{refinement} $P'$ of $P$, which modifies $P$ by filling a single hole in $P$ with an expression. The user either accepts the refinement if it matches the ``true'' program that the user is aiming to write (in which case we update $P\gets P'$), or rejects it otherwise (in which case a different refinement is proposed). This process continues until $P$ is concrete (i.e., it has no holes), at which point our algorithm returns $P$.

\paragraph{Selecting a candidate refinement.}

The key step performed by our algorithm on each iteration is to select a candidate refinement $P'$ of $P$ with which to question the user. In Figure~\ref{fig:ex}, an example of a refinement is shown on the third line. This refinement is constructed from the sketch on the first line using the productions shown on the second line, which says that the hole named \texttt{t} should be replaced with the expression
\begin{align*}
\texttt{authors INNER-JOIN ??t\_new:table ON ...}
\end{align*}
Note that this expression contains new holes; if the user accepts this refinement, then our algorithm will need to fill in these holes on subsequent iterations.

Intuitively, our goal is to choose the question that elicits the most information about the true program---i.e., the question should cut down the search space by as much as possible. To formalize this intuition, we use the user-provided sketch to induce a probability distribution over completions of that sketch. Then, we can ask for the refinement that prunes the largest number of completions in expectation, where each completion is weighted by its probability.
\footnote{An optimal approach would be to optimize this objective over a sequence of questions; however, this approach quickly becomes intractable. In fact, the performance of the greedy strategy (in terms of expected number of questions used) is a log-factor of the performance of the optimal approach~\cite{dasgupta2005analysis}.}
Computing this refinement is challenging due to the exponential size of the search space over programs. Instead, we use MCMC to randomly sample a finite number of completions $\complete{P}$ of $P$, and then choose the refinement $P'$ of $P$ according to an estimate of the above objective using these samples.

\paragraph{User interaction.}

Once our algorithm has selected a refinement $P'$, it shows this refinement to the user. The user must either accept $P'$ if the true program is a completion of $P'$, or reject it otherwise. In our example in Figure~\ref{fig:ex}, the true program on the last line is a refinement of the refinement on the third line; thus, the user accepts this refinement; then, $P$ is updated to be $P'$, and the interactive process continues.

A key constraint is that we need to choose the space of possible refinements to ensure that users can understand whether the refinement matches the true program. As discussed above, we assume that users are familiar with SQL, and the key challenge is making sure they can understand whether the tables and columns in the database are the right ones to use in various parts of the query.

In particular, as discussed above, the primary purpose of our algorithm is to determine the sequence of joins that are needed to construct the flat table that the user needs to perform subsequent tasks. In particular, there are two kinds of refinements considered by our algorithm:
(i) specifying an inner-join expression of the form
\begin{align*}
\texttt{??}\rightarrow t~\texttt{INNER-JOIN ?? ON ?? = ??}~
\end{align*}
or a single table to use---i.e., $\texttt{??}\rightarrow t$, in which case a summary of $t$ is shown to the user, or (ii) specifying which column to use in a select, project, or inner-join operation---i.e., $\texttt{??}\Rightarrow c$, in which case a summary of $c$ is shown to the user.

In our example in Figure~\ref{fig:ex}, the refinement on the third line is obtained from the sketch by filling the hole named \texttt{t} with the table \texttt{authors} (inner-joined with another, currently unspecified table). Thus, our algorithm would show a summary of the authors table in Figure~\ref{fig:exdb} to the user (e.g., the first few rows of this table). This information suffices for the user to decide whether to accept the refinement, since they see that it includes author names that want included in the flat table constructed by the true program.

If the user accepts, we update $P$ to equal $P'$. We continue the iterative process until $P$ is concrete, at which point it returns $P$. Because we question the user at every step (including the final program $P$), we guarantee that the final program is the one desired by the user.

\section{Sketch Language}
\label{sec:prob}

\begin{figure}
\begin{minipage}{0.32\textwidth}
\begin{align*}
Q~::=~&\Pi_{C,...,C}(S)~\{\Phi\} \\
S~::=~&\sigma_{\Psi}(I)~\{\Phi\} \\
I~::=~&T~\{\Phi\}\mid T\Join_{C,C}I~\{\Phi\} \\
\Psi~::=~&\true\mid C~R~V\mid\Psi\wedge\Psi\mid\Psi\vee\Psi \\
R~::=~&\le~\mid~<~\mid~=~\mid~>~\mid~\ge
\end{align*}
\end{minipage}
\begin{minipage}{0.1\textwidth}
\begin{align*}
T~::=~&t_1\mid...\mid t_n \\
C~::=~&c_1\mid...\mid c_m \\
X~::=~&x_1\mid...\mid x_{\ell}
\end{align*}
\end{minipage}
\begin{minipage}{0.48\textwidth}
\begin{align*}
\Phi~::=~&\true\mid X\in C\mid\textsf{contains}(c,r)\mid C~U~X\mid\Phi\wedge\Phi \\
U~::=~&\lesssim\;\mid\;\simeq\;\mid\;\gtrsim
\end{align*}
\end{minipage}
\caption{Syntax of database queries, with start symbol $Q$. Here, $t_1,...,t_n$ are tables, $c_1,...,c_m$ are columns, and $x_1,...,x_{\ell}$ are constants (i.e., integers, floats, strings, and regular expressions). The language is based on~\cite{yaghmazadeh2017sqlizer}, except there are no aggregation operations, and queries are normalized so projection and selection operations are executed last. We permit two kinds of holes: (i) a table in a sequence of inner-join operations (corresponding to nonterminal $I$), and (ii) columns in any operation (corresponding to nonterminal $C$).}
\label{fig:querysyntax}
\end{figure}

We consider a domain-specific language (DSL) $\D$ of database queries based on a fragment of SQL that only includes select, project, and inner-join operations. Its syntax is a context-free grammar $\GG=(V,\Sigma,\mathcal{R},Q)$ with non-terminals $V$, terminals $\Sigma$, productions $\mathcal{R}$, and start symbol $Q$. This grammar is shown in Figure~\ref{fig:querysyntax}. Projection of a table $t$ onto a list of columns $c_1,...c_n$ is denoted $\Pi_{c_1,...,c_n}(t)$, selection of rows that satisfy a predicate $\psi$ from table $t$ is denoted $\sigma_{\psi}(t)$, and the inner-join of tables $t_1$ and $t_2$ on column $c_1$ in $t_1$ and $c_2$ in $t_2$ is denoted $t_1\Join_{c_1,c_2}t_2$. The semantics $\semantics{\cdot}:\LL(\GG)\to\T$ maps programs $P\in\LL(\GG)$ to tables $t\in\T$. They ignore the soft constraints $\Phi$; otherwise, they are standard, so we omit them.

Note that we have constrained to expressions of the form
\begin{align*}
\Pi_{C,...,C}(\sigma_{\Psi}(T\Join_{C,C}\cdots\Join_{C,C}T)).
\end{align*}
In general, by using the relational algebra, any composition of select, project, and inner-join operations can be equivalently expressed in this form. Since our focus is on the sequence of inner-join operations, we assume that the user will specify the structure of the project and select operations, and only leave tables in the inner-join operation on the inside as holes (columns can be left as holes anywhere in the query). Note that this grammar also includes soft constraints $\phi$ on tables $t$ (denoted $t~\{\phi\}$), which we discuss below.

\paragraph{Notation.}

We establish some standard notation. Consider a sequence $\alpha=A_1...A_k\in(V\cup\Sigma)^*$. Suppose that $A_i\in V$; then, we can apply a production $A_i\to A_{i1}...A_{ih}$ to obtain
\begin{align*}
\alpha'=A_1...A_{i-1}A_{i1}...A_{ih}A_{i+1}...A_k,
\end{align*}
and we denote this relationship by $\alpha\Rightarrow\alpha'$. Furthermore, if there exists a sequence $\alpha\Rightarrow\alpha'\Rightarrow...\Rightarrow\alpha''$, then we say $\alpha''$ can be \emph{derived} from $\alpha$, which we denote by $\alpha\xRightarrow{*}\alpha''$.

We refer to a sequence $\alpha$ such that $A\xRightarrow{*}\alpha$ for some nonterminal $A\in V$ as an \emph{expression}. We let $\LL(\GG,A)$ denote the \emph{concrete} expressions $\alpha\in\Sigma^*$ that can be derived from $A$---i.e., $A\xRightarrow{*}\alpha$. Note that $\LL(\GG,Q)=\LL(\GG)$---i.e., the space of programs defined by the grammar $\GG$ is the set of expressions that can be derived from the start symbol $Q$.

\paragraph{Sketches.}

Our algorithm keeps track of programs that have holes. In particular, a \emph{sketch}~\cite{solar2005programming,solar2006combinatorial} is a a sequence $P\in(V\cup\Sigma)^*$ such that $P$ can be derived from $Q$---i.e., $Q\xRightarrow{*}P$. We refer to a nonterminal in $P$ as a \emph{hole}. We restrict to holes $A$ that are either $A=I$ (i.e., tables in the sequence of inner-join operations) or $A=C$ (i.e., columns). We associate a name $s$ (i.e., a string) with each hole in $P$; a name identifies different holes that should be filled using identical expressions. A sketch is \emph{complete} (also called a \emph{concrete program}) if $P\in\Sigma^*$---i.e., it has no holes (which implies that $P\in\LL(\GG)$). In our language, the sketch on the first line of Figure~\ref{fig:ex} is written
\begin{align*}
P_{\text{author}}&=\Pi_{C:\texttt{c\_name}}\left(\sigma_{C:\texttt{c\_year}=1948}\left(I:\texttt{t}~\{\phi\}\right)\right) \\
\phi&=(\textsf{contains}(C:\texttt{c\_name},~\text{``.*Church.*''}) \\
&\qquad\wedge(C:1900\lesssim\texttt{c\_year}\lesssim 2020).
\end{align*}
We have dropped the types from holes $A$, since they are determined by the value of the hole---i.e., $\texttt{table}$ if $A=I$ or $\texttt{column}$ if $A=C$. Instead, we have used the notation $A:s$ to denote the hole $A$ with associated name $s$. We have also dropped soft constraints $\{\phi\}$ when $\phi=\true$.

\paragraph{Abstract syntax trees.}

Internally, our algorithm represents a sketch $P$ using its abstract syntax tree (AST), which is a representation of the derivation of $P$ in $\GG$. For convenience, we use $P$ to denote both the sequence $P\in(V\cup\Sigma)^*$ as well as its AST. We denote the nodes of $P$ by $\nodes{P}$, the internal nodes by $\internal{P}$, and the leaves by $\leaves{P}$.

Each node $N$ in $P$ is associated with a symbol $A_N\in V\cup\Sigma$, which is the symbol associated with $N$ in the derivation of $P$. An internal node is always labeled with a nonterminal. If $P$ is complete, then each leaf node of $P$ is labeled with a terminal; otherwise, a leaf node of $P$ may be labeled with either a terminal or a nonterminal. Note that holes correspond to leaf nodes of $P$ labeled with a nonterminal; we denote the set of holes by $\holes{P}\subseteq\leaves{P}$.

Finally, we use $\alpha_N$ to denote the subtree of $P$ at $N$. Note that $\alpha_N$ can also be thought of as a subexpression $\alpha_N\in(\Sigma\cup V)^*$ in $P$ derived from $A_N$---i.e., $A_N\xRightarrow{*}\alpha_N$.

As an example, $P_{\text{author}}$ corresponds to the AST
\begin{align*}
\begin{tikzpicture}
[sibling distance=4em,
level distance=2.5em,
label distance=-2pt,
font=\scriptsize,
MyCircle/.style={
    draw, circle, align=center, top color=white, bottom color=white,
    prefix after command= {\pgfextra{\tikzset{every label/.style={black}}}}
}]
\node[MyCircle,label={180:{$Q$}}]{$N_1$}
child {
    node[MyCircle,label={180:{$\Pi$}}]{$N_2$}
    edge from parent[->,>=latex]
}
child {
    node[MyCircle,label={180:{$C$}}]{$N_3$}
    edge from parent[->,>=latex]
}
child {
    node[MyCircle,label={180:{$S$}}]{$N_4$}
    edge from parent[->,>=latex]
    child {
        node[MyCircle,label={180:{$\sigma$}}]{$N_5$}
        edge from parent[->,>=latex]
    }
    child {
        node[MyCircle,label={180:{$\Psi$}}]{$N_6$}
        edge from parent[->,>=latex]
        child {
            node[MyCircle,label={180:{$C$}}]{$N_7$}
            edge from parent[->,>=latex]
        }
        child {
            node[MyCircle,label={180:{$=$}}]{$N_8$}
            edge from parent[->,>=latex]
        }
        child {
            node[MyCircle,label={180:{$1948$}}]{$N_9$}
            edge from parent[->,>=latex]
        }
    }
    child {
        node[MyCircle,label={180:{$I$}}]{$N_{10}$}
        edge from parent[->,>=latex]
    }
    child {
        node[MyCircle,label={180:{$\Phi$}}]{$N_{11}$}
        edge from parent[->,>=latex]
        child {
            node[MyCircle]{$...$}
            edge from parent[->,>=latex]
        }
    }
};
\end{tikzpicture}
\end{align*}
where we have omitted the subtree rooted at the child of $N_{11}$. Each node $N$ in the AST is labeled with its corresponding symbol $A_N$. The holes shown are $N_3$ (named $\texttt{c\_name}$), $N_7$ (named $\texttt{c\_year}$) and $N_{10}$ (named $\texttt{t}$) (there are additional holes not shown). An example of a subexpression $\alpha_N$ is
\begin{align*}
\alpha_{N_5}=\sigma_{C:\texttt{c\_year}=1948}(I:\texttt{t}~\{\phi\}).
\end{align*}

\paragraph{Refinements.}

One sketch $P'$ is a \emph{refinement} of another one $P$ if $P\xRightarrow{*}P'$---i.e., $P'$ can be obtained from $P$ by filling in the holes of $P$ with expressions in $\GG$ (which may contain additional holes). Note that the nodes of $P$ are a subset of the nodes of $P'$; we use $\iota:\nodes{P}\to\nodes{P'}$ to denote the natural injection from nodes of $P$ to nodes of $P'$.

Furthermore, for a hole $H\in\holes{P}$, we say $H$ is \emph{filled} with the expression $\alpha_{\iota(H)}$, where $\alpha_{\iota(H)}$ is the subexpression of $P'$ at $\iota(H)$. Note that conversely, given a set of expressions $\{\alpha_H\mid H\in\holes{P}\}$ such that $A_H\xRightarrow{*}\alpha_H$, we can construct a refinement $P'$ of $P$ by replacing $A_H$ with $\alpha_H$ in $P$.

In our example in Figure~\ref{fig:ex}, the sketch on the third line is obtained by applying the sequence of productions
\begin{align*}
I&\Rightarrow T\Join_{C:\texttt{c\_new0},C:\texttt{c\_new1}}I:\texttt{t\_new} \\
&\Rightarrow\texttt{authors}\Join_{C:\texttt{c\_new0},C:\texttt{c\_new1}}I:\texttt{t\_new}
\end{align*}
to $P_{\text{author}}$. In particular, we then obtain the sketch
\begin{align*}
P_{\text{author}}'&=\Pi_{C:\texttt{c\_name}}\left(\sigma_{C:\texttt{c\_year}=1948}\left(t~\{\phi\}\right)\right) \\
t&=\texttt{authors}\Join_{C:\texttt{c\_new0},C:\texttt{c\_new1}}I:\texttt{t\_new}
\end{align*}
to $P_{\text{author}}$---i.e., $P_{\text{author}}\xRightarrow{*}P_{\text{author}}'$. Note that $P_{\text{author}}'$
has three new holes \texttt{c\_new0}, \texttt{c\_new1}, and \texttt{t\_new} compared to $P_{\text{author}}$. Furthermore, note that $P_{\text{author}}$ is obtained by filling the hole $I:\texttt{t}$ in $P_{\text{author}}$ with the expression $t$.

\paragraph{Completions.}

Our algorithm assumes that the true program can be derived from the sketch provided by the user. In particular, a \emph{completion} $\complete{P}$ of $P$ is a complete refinement of $P$---i.e., a refinement of $P$ that has no holes; we denote the set of completions of $P$ by $\PP_P$. In this case, the subexpression $\alpha_{\iota(H)}$ used to fill hole $H\in\holes{P}$ is concrete (i.e., it has no holes). Thus, we have $\alpha_{\iota(H)}\in\LL(\GG,A_H)$.

In our example in Figure~\ref{fig:ex}, the last line shows a completed sketch, which can equivalently be written
\begin{align*}
\complete{P}_{\text{author}}&=\Pi_{\texttt{name}}\left(\sigma_{\texttt{year}=1948}\left(\complete{t}~\{\complete{\phi}\}\right)\right) \\
\complete{t}&=\texttt{authors}\Join_{\texttt{aid},\texttt{aid}}\texttt{writes}\Join_{\texttt{pid},\texttt{pid}}\texttt{publications} \\
\complete{\phi}&=(\textsf{contains}(\texttt{name},~\text{``.*Church.*''}) \\
&\qquad\wedge(\texttt{year}\ge1900)\wedge(\texttt{year}\le2020).
\end{align*}
This sketch is a completion of $P_{\text{author}}$ (and of $P_{\text{author}}'$). Note that $\complete{P}_{\text{author}}$ can be obtained from $P_{\text{author}}$ by filling hole \texttt{t} with $\complete{t}$, hole \texttt{c\_name} with \texttt{name}, and hole \texttt{c\_year} with \texttt{year}.

\begin{figure}
\begin{align*}
\semantics{\Pi_{c_1,...,c_k}(s)~\{\phi\}}_{\varphi}&=\semantics{\phi}_{\varphi}~\semantics{\Pi_{c_1,...,c_k}(s)}+\semantics{s}_{\varphi} \\
\semantics{\sigma_{\psi}(i)~\{\phi\}}_{\varphi}&=\semantics{\phi}_{\varphi}~\semantics{\sigma_{\psi}(i)}+\semantics{i}_{\varphi} \\
\semantics{i~\{\phi\}}_{\varphi}&=\semantics{\phi}_{\varphi}~\semantics{i} \\
\semantics{t\Join_{c,c'}i~\{\phi\}}_{\varphi}&=\semantics{\phi}_{\varphi}~\semantics{t\Join_{c,c'}i}+\semantics{i}_{\varphi}
\end{align*}
\begin{align*}
\semantics{\true}_{\varphi}&=\lambda t\;.\;0 \\
\semantics{x\in c}_{\varphi}&=\lambda t\;.\;\mathbb{I}\left[x\in t[c]\right]_{\varphi} \\
\semantics{\textsf{contains}(c,r)}_{\varphi}&=\lambda t\;.\;\mathbb{I}\left[\exists x\in t[c]~.~\textsf{match}(x,r)]\right]_{\varphi} \\
\semantics{c~u~x}_{\varphi}&=\lambda t\;.\;\frac{1}{|t[c]|}\sum_{x'\in t[c]}\mathbb{I}\left[x'~\semantics{u}~x\right] \\
\semantics{\phi\wedge\phi'}_{\varphi}&=\lambda t\;.\;\semantics{\phi}_{\varphi}~t+\semantics{\phi'}_{\varphi}~t
\end{align*}
\caption{Semantics of our soft constraints. The first four lines are the semantics for table expressions $t~\{\phi\}\in\LL(\GG,A)$ for $A\in\{Q,S,I\}$; in this case, $\semantics{\cdot}_{\varphi}\in\mathbb{R}$. The last four lines are the semantics for soft constraints $\phi\in\LL(\GG,\Phi)$; in this case, $\semantics{\cdot}_{\varphi}\in\{\T\to\mathbb{R}\}$ is a mapping from tables to the reals. For a table $t\in\T$ and a column $c$, $t[c]$ is the list of values $x$ in column $c$ of $t$. For $\semantics{u}$, $\semantics{\lesssim}$ is $\le$, $\semantics{\simeq}$ is $=$, and $\semantics{\gtrsim}$ is $\ge$. We use $\mathbb{I}$ to denote the indicator function, and $\textsf{match}(x,r)$ to denote that string $x$ matches regular expression $r$.}
\label{fig:soft}
\end{figure}

\paragraph{Soft constraints.}

Our language includes constraints of the form $\alpha\sim\{\phi\}$, where $\alpha\in\LL(\GG,A)$ for some $A\in\{Q,S,I\}$ is a table expression (i.e., a select, project, or inner-join operation), and $\phi\in\LL(\GG,\Phi)$ is a \emph{soft constraint} on tables $t\in\T$.

The DSL semantics $\semantics{\cdot}$ ignore these specifications. Instead, we define an additional semantics $\semantics{\cdot}_{\varphi}:\LL(\GG)\to\mathbb{R}$; these semantics are shown in Figure~\ref{fig:soft}. In particular, $\semantics{P}_{\varphi}$ can be interpreted as a score encoding how well $P$ satisfies the soft constraints; a high score means that $P$ is satisfies the constraints very well. Furthermore, soft constraints encode user expectations, so a high score means that $P$ is a close match for what the user is expecting.

The soft constraints are on the values obtained when evaluating $P$. Intuitively, these semantics interpret $\phi$ as a soft constraint on the value $\semantics{\alpha}$ obtained by evaluating the expression $\alpha$ preceding $\phi$ using the DSL semantics $\semantics{\cdot}$. In particular, $\semantics{\phi}_{\varphi}:\T\to\mathbb{R}$ is a mapping from tables to real numbers. Then, the semantics for expressions of the form $\alpha~\{\phi\}$ are obtained by applying $\phi$ to $\alpha$ to obtain $\semantics{\phi}_{\varphi}~\semantics{\alpha}\in\mathbb{R}$ (since $\semantics{\alpha}\in\T$). In addition, these semantics aggregate the values obtained from additional specifications in the expression $\alpha$ by summing them together. In Figure~\ref{fig:soft}, the first four lines show the semantics for expressions $\alpha~\{\phi\}$, and the last four lines show the semantics for soft constraints $\phi$.

The semantics for $\phi=\true$ always evaluates to $0$; this choice is since $\true$ is the default specification, and evaluating to $0$ ensures that these specifications do not affect the score $\semantics{P}_{\varphi}$. For containment $x\in c$, the score is $1.0$ if the value $x$ is in column $c$ of the given table $t$, and $0.0$ otherwise. For an expression $c\lesssim x$, where $x$ is a value and $c$ is a column, the score is the fraction of values $x'$ in column $c$ of the table $t$ that satisfies $x'\le x$; $\simeq$ and $\gtrsim$ are similar. In all these cases, $x$ can be a integer, float, string, or regular expressions. For strings, inequalities are interpreted as lexicographical ordering. For regular expressions, we restrict to containment and approximate equality $\simeq$; they are interpreted as typical regular expression matching (e.g., for containment, there must exist some $x'\in t[c]$ such that $x'\in\LL(x)$). For conjunctions $\phi\wedge\phi'$, we add up the score based on $\phi$ and the one based on $\phi'$. We make this choice since we typically use conjunctions to place multiple unrelated constraints on a table.

As an example, for the program $\complete{P}_{\text{author}}$, there is a single soft constraint $\complete{\phi}$. This constraint applies to the table $\complete{t}$, which evaluates to the table shown in Figure~\ref{fig:exdbex}. For this table, the soft constraint $\texttt{name}\simeq2$ evaluates to $0.33$, $\texttt{year}\gtrsim1900$ to $1.0$, and $\texttt{year}\lesssim2020$ to $1.0$; thus, $\semantics{\complete{P}_{\text{author}}}=2.33$.

\section{Interactive Synthesis Algorithm}
\label{sec:alg}

\begin{algorithm}[t]
\begin{algorithmic}
\Procedure{InteractiveSynthesize}{Sketch $P$}
\State $\N\gets\varnothing$
\While{$\neg\textsc{IsComplete}(P)$}
\State $\Q_P\gets\textsc{ConstructCandidateQueries}(P)\setminus\N$
\State $\PC\gets\textsc{SampleCompletions}(\pi_{P,\mathcal{N}})$
\State $\hat{Q}\gets\operatorname*{\arg\max }_{Q\in\Q_P}\scorehat{Q;\PP}$
\State $b\gets\OO(\hat{Q})$
\State \textbf{if} $b$ \textbf{then} $P\gets\hat{Q}$ \textbf{else} $\mathcal{N}\gets\mathcal{N}\cup\{\hat{Q}\}$
\EndWhile
\State \Return $P$
\EndProcedure
\end{algorithmic}
\caption{Our interactive synthesis algorithm.}
\label{alg:synth}
\end{algorithm}

Our algorithm takes as input a sketch $P$ and returns a completion $\complete{P}$ of $P$ that satisfies the user intent. At a high level, it iteratively refines $P$, questioning the user at step to ensure that the refinement satisfies the user's intent. Our algorithm returns $P$ once it is complete. Assuming the user answers correctly, it is guaranteed to return the true program after a number of iterations that is polynomial
in the size of $\complete{P}$. Each iteration of our algorithm computes a question $\hat{Q}$ (i.e., a candidate refinement of the current sketch $P$) in two steps:
\begin{enumerate}
\item Compute a set of candidate questions $\Q$.
\item Compute the refinement $\hat{Q}\in\Q$ that maximizes the expected reduction in the size of the search space.
\end{enumerate}
Then, our algorithm questions the user on the candidate refinement $\hat{Q}$. If the user accepts $\hat{Q}$, then our algorithm updates $P\gets\hat{Q}$; otherwise, $\hat{Q}$ is added to a set of negative responses $\N$ that are avoided in subsequent iterations. By doing so, our algorithm maintains the invariant that the true program $\complete{P}^*$ is a completion of the current sketch $P$. In particular, this invariant is preserved assuming the user provides a valid initial sketch and answers correctly at each iteration.

We describe our algorithm in more detail below. We initially ignore the impact of negative user responses, and then describe how to handle them in Section~\ref{sec:neg}.

\subsection{Constructing Candidate Questions}

A \emph{question} $Q$ is a sketch that is a refinement of the current sketch $P$. We construct the \emph{candidate questions} $\Q_P$ as follows:
\begin{itemize}
\item For each column hole $C$ in $P$, we include the sketch obtained by filling $C$ using a production $C\Rightarrow c_i$ for some column $c_i$.
\item For each table hole $I$, we include the sketches obtained by filling $I$ using one of the sequences of productions
\begin{align*}
&I\Rightarrow T\Rightarrow t_i \quad \text{or} \quad
I\Rightarrow T\Join_{C,C}I\Rightarrow t_i\Join_{C,C}I
\end{align*}
for some table $t_i\in\T$. In both cases, we implicitly use the default specification $\Phi\Rightarrow\true$.
\end{itemize}
Note that all new holes correspond to nonterminal $I$ or $C$.

Furthermore, we ensure that holes with the same name are filled using the same expressions. First, we include names on new holes in the productions used to fill holes---i.e.,
\begin{align*}
I\xRightarrow{*}t_i\Join_{C:s_1,C:s_2}I:s_3.
\end{align*}
Second, if we replace hole $A:s$, then we also replace all other holes named $s$ using the same sequence of productions.

As an example, $\Q_{P_{\text{author}}}$ includes the sketch
\begin{align*}
P_{\text{author}}''&=\Pi_{\texttt{name}}\left(\sigma_{C:\texttt{c\_year}=1948}\left(I:\texttt{t}~\{\phi\}\right)\right) \\
\phi&=(\textsf{contains}(C:\texttt{c\_name},~\text{``.*Church.*''}) \\
&\qquad\wedge(C:1900\lesssim\texttt{c\_year}\lesssim 2020).
\end{align*}
where both holes $C:\texttt{c\_name}$ have been filled using $\texttt{name}$; $\Q_{P_{\text{author}}}$ also includes $P_{\text{author}}'$, among others.

\subsection{Computing Good Questions}

Our algorithm selects a refinement $\hat Q\in\Q_P$ on which to question the user using a greedy active learning strategy~\cite{dasgupta2005analysis} where we use sampling to estimate the score.

\paragraph{User responses.}

Our goal is to choose questions that cut off as much of the search space as possible. To do so, we need to define what part of the search space is cut off by a given user response. To this end, we represent the user as an oracle $\OO:\Q\to\bool$, where $\Q$ is the set of all possible questions and $\bool=\{\true,\false\}$. We assume the user response $\OO(Q)$ indicates whether the true program $\complete{P}^*$ (i.e., the program the user desires) is a completion of the question $Q$---i.e., $\OO(Q)=\mathbb{I}[Q\xRightarrow{*}\complete{P}^*]$, where $\mathbb{I}$ is the indicator function. For example, $P_{\text{author}}'$ is a candidate question for $P_{\text{author}}$; furthermore, assuming the true program is $\complete{P}_{\text{author}}$, then $\OO(P_{\text{author}}')=\true$ since $\complete{P}_{\text{author}}$ is a completion of $P_{\text{author}}'$.

Note that the search space is the set $\PP_P$ of completions of the current sketch $P$. If the user responds $\OO(Q)=\true$, we remove completions $\complete{P}\in\PP_P$ such that $\complete{P}\not\in\PP_Q$---i.e., programs that are not completions of $Q$. Conversely, if $\OO(Q)=\false$, we remove completions $\complete{P}\in\PP_P$ such that $\complete{P}\in\PP_Q$.

\paragraph{Scoring questions.}

We score candidate questions $Q\in\Q_P$ based on the expected fraction of the search space that they cut off. To formalize this notion, we assign probabilities to completions $\complete{P}\in\PP_P$ of the current sketch $P$ using the scores $\semantics{\complete{P}}_{\varphi}$ based on the soft constraints provided by the user. In particular, we define a probability distribution $\pi_P$ over the completions $\complete{P}$ of $P$ as follows:
\begin{align*}
\pi_P(\complete{P})=\frac{1}{Z_P}e^{\semantics{\complete{P}}_{\varphi}}
\hspace{0.1in}\text{where}\hspace{0.1in}
Z_P=\sum_{\complete{P}'}e^{\semantics{\complete{P}'\in\PP_P}_{\varphi}}
\end{align*}
where $\PP_P$ denotes all completions of $P$. In particular, completions of $P$ with higher score have higher probability. Thus, programs that are more likely according to the user-provided soft constraints have higher probability.

Then, we score a candidate question based on the fraction of the search space $\PP_P$ of completions $\complete{P}$ of $P$ remaining based on the user's response, weighted by the probabilities $\pi_P(\complete{P})$. We weight according to these probabilities to focus on disambiguating among programs that are likely to be the true program $\complete{P}^*$. For a question $Q\in\Q_P$, if the user accepts $Q$, then the fraction of the search space remaining is
\begin{align*}
\pi_+=\mathbb{P}_{\complete{P}\sim\pi_P}[Q\xRightarrow{*}\complete{P}]=\sum_{\complete{P}}\mathbb{I}[Q\xRightarrow{*}\complete{P}]\cdot\pi_P(\complete{P}).
\end{align*}
On the other hand, if the user rejects $Q$, then the fraction of the search space remaining is $\pi_-=1-\pi_+$.

However, we cannot know the fraction of the search space that is cut off by a question $Q$ without knowing the user response $\OO(Q)$. To address this issue, we interpret $\pi_P(\complete{P})$ as the probability that the (unknown) true completion $\complete{P}^*$ is $\complete{P}$. Then, we can compute the probability that the user responds $\true$ or $\false$---in particular, the probability that the user responds $\true$ is exactly $\pi_+$, since $\pi_+$ is the probability that $\complete{P}^*$ is a refinement of $Q$. Analogously, $\pi_-$ is the probability that the user responds $\false$. Thus, the expected score is
\begin{align*}
\score{Q;\pi_P}=&\sum_{b\in\mathbb{B}}(\text{search space pruned if }\OO(Q)=b) \\
&\hspace{0.3in}\times(\text{probability that }\OO(Q)=b) \\
=&\pi_+\cdot\pi_-+\pi_-\cdot\pi_+ \\
=&2\pi_+\cdot(1-\pi_+).
\end{align*}
The ideal question is one where $\pi_+=\pi_-$---then, no matter how the user responds, it cuts the search space in half.

\paragraph{Computing good completions.}

Note that the score of a question is the expected fraction of the search space pruned if we ask the user that question. Thus, we would ideally choose the completion that maximizes the score
\begin{align*}
Q^*=\operatorname*{\arg\max}_{Q\in\Q_P}\score{Q;\pi_P}.
\end{align*}
However, it is intractable to compute the score exactly due to the sum over completions $\complete{P}\in\PP_P$. Instead, we use random samples $\complete{P}\sim\pi_P$ to approximate the score---i.e., given a set $\PP\subseteq\PP_P$ of i.i.d. samples from $\pi_P$, we use the approximation
\begin{align*}
\score{Q;\pi_P}\approx\scorehat{Q;\PP}=2\cdot\hat{\pi}_+\cdot(1-\hat{\pi}_+),
\end{align*}
where $\hat{\pi}_+=\frac{1}{|\PP|}\sum_{\complete{P}\in\PP_P}\mathbb{I}[Q\xrightarrow{*}\complete{P}^*]$. Then, we choose question
\begin{align*}
\hat{Q}=\operatorname*{\arg\max}_{Q\in\Q_P}\scorehat{Q;\PP}.
\end{align*}

\paragraph{Sampling completions.}

A remaining challenge is how to sample completions $\complete{P}\sim\pi_{P_\phi}$. The difficulty is that we can only compute the unnormalized probabilities $\pi_{P_\phi}(\complete{P})$. We use a standard approach to randomly sample completions---namely, the Metropolis-Hastings (MH) algorithm~\cite{chib1995understanding,schkufza2013stochastic}. We give a brief overview here. To use this algorithm, we must define (i) a way to sample an initial completion, and (ii) a way to sample a neighbor $\complete{P}'$ of a completion $\complete{P}$.

To sample an initial completion, we independently sample an expression to fill each hole $H\in\holes{P}$ of the current sketch $P$. In particular, for each hole $H$, we sample a random expression $\alpha_H\sim\LL(\GG,A_H)$, and then use $\alpha_H$ to fill $H$. Once all the holes have been filled, we obtain a completion $\complete{P}$. Here, we think of $\LL(\GG,A_H)$ as a probabilistic grammar in a standard way---i.e., by using the uniform distribution over productions for each nonterminal. Next, to sample a neighbor $\complete{P}'$ of a completion $\complete{P}$, we uniformly randomly choose a single hole $H\in\holes{P}$, and replace the expression $\alpha_H$ in $\complete{P}$ with a newly sampled expression $\alpha_H'\sim\LL(\GG,A_H)$. This produces a modified completion $\complete{P}'$.

Then, MH starts by sampling an initial completion $\complete{P}$. Then, for a fixed number of steps, it samples a neighbor $\complete{P}'$ of $\complete{P}$; if the (unnormalized) probability $\pi_P(\complete{P}')$ is larger than $\pi_P(\complete{P})$ (i.e., $\complete{P}'$ better matches the soft constraints than $\complete{P}$), then we update $\complete{P}\gets\complete{P}'$. Otherwise, we still perform this update with some probability; this probability is computed to ensure that asymptotically, $\complete{P}$ is a random sample from $\pi_P$.

\subsection{Handling Negative Responses}
\label{sec:neg}

So far, we have ignored the impact of user responses $\OO(Q)=\false$ on our search space. If a user responds $\OO(Q)$, then the current sketch $P$ does not change; instead, completions that do not match $Q$ are removed from our search space. In particular, our algorithm keeps track of questions $Q\in\N$ for which $\OO(Q)=\false$. Then, our search space is actually
\begin{align*}
\PP_{P,\N}=\{\complete{P}\in\PP_P\mid\forall Q\in\N~.~Q\centernot{\xRightarrow{*}}\complete{P}\}.
\end{align*}
In other words, $\PP_{P,\N}$ omits completions that match any of the questions $Q\in\N$. Then, we modify $\pi_P$ to take this restriction into account---i.e., we define the distribution
$\pi_{P,\N}(\complete{P})\propto\pi_P(\complete{P})\cdot\mathbb{I}[\complete{P}\in\PP_{P,\N}]$
over completions $\complete{P}$ of $P$. In other words, $\pi_{P,\N}$ is $\pi_P$ conditioned on the event that $\complete{P}\in\PP_{P,\N}$.

Then, our algorithm remains the same, except we use $\pi_{P,\N}$ in place of $\pi_P$ when sampling completions $\PP$. We use rejection sampling to sample $\complete{P}\sim\pi_{P,\N}$---i.e., we repeatedly obtain samples $\complete{P}\sim\pi_P$ until we find one that satisfies the condition $\mathbb{I}[\complete{P}\in\PP_{P,\N}]$. To check this condition, we simply iterate over each $Q\in\N$ and check if $\complete{P}$ is a completion of $Q$.

\subsection{Overall Algorithm}

Our algorithm is shown in Algorithm~\ref{alg:synth}. It first constructs the set $\Q_P$ of candidate questions; in this step, it also removes questions $Q\in\N$ for which the user has already responded negatively. Next, it samples an i.i.d. set of completions $\PP$ from $\pi_{P,\N}$. Then, it chooses the best completion $\hat Q$ based on these samples. Finally, it questions the user on $\hat{Q}$. If $\OO(\hat Q)=\true$, then it updates $P\gets\hat{P}$; otherwise, it adds $\hat{Q}$ to the questions $\N$ with negative responses. Finally, it iteratively continues this process until $P$ is complete, at which point it returns $P$.

We prove that assuming the user provides a valid initial sketch and responds correctly, then our algorithm returns $\complete{P}^*$. We emphasize that our key contributions are our design decisions---i.e., the kind of input we require from the user. Our theoretical guarantees follow straightforwardly given these choices; wes give proofs in Appendix~\ref{sec:appproofs}. First, we prove that if our algorithm returns, then its return value is correct.
\begin{theorem}
\label{thm:main}
Suppose (i) the initial sketch $P$ provided by the user satisfies $P\xRightarrow{*}\complete{P}^*$, and (ii) the user responses are $\OO(Q)=\mathbb{I}[Q\xRightarrow{*}\complete{P}^*]$. Then, if our algorithm terminates, it returns $\complete{P}^*$.
\end{theorem}
Next, we prove that our algorithm is guaranteed to terminate. In fact, we prove that it is guaranteed to do so in a polynomial number of iterations.
\begin{theorem}
\label{thm:complete}
Our algorithm terminates after $O((n+m)\cdot|\complete{P}^*|^2)$ iterations, where $|\complete{P}^*|$ is the number of nodes in the AST of $\complete{P}^*$, $n$ is the number of tables in the database, and $m$ is the number of columns in the database.
\end{theorem}
This bound holds regardless of how our algorithm chooses questions. While polynomial, the number of iterations can still be large if the questions are chosen poorly---thus, in practice, choosing good questions is important.

\section{Implementation}
\label{sec:impl}

We have implemented our algorithm in a tool called \toolname. We briefly discuss our implementation, and give details in Appendix~\ref{sec:appimpl}. First, our implementation restricts the search space of complete programs---i.e., by restricting to inner-join operations on columns with matching keys and by imposing type constraints; by doing so, we reduce the number of iterations needed. We also modify the candidate questions for inner-join operations to fill columns involved with a join at the same time as a table---i.e., $t_i\Join_{C,C}I\xRightarrow{*}t_i\Join_{c,c'}t_j$ and $t_i\Join_{C,C}I\xRightarrow{*}t_i\Join_{c,c'}t_j\Join_{C,C}I$. Finally, we precompute the soft constraints so we do not have to evaluate database queries during the execution of our algorithm; this change slightly modifies the semantics $\semantics{\cdot}_{\varphi}$.

\section{Evaluation}
\label{sec:eval}

We evaluate our approach on a database of academic publications~\cite{li2014constructing}. This database has 16 tables; it comes with a dataset of 196 examples of SQL programs described in natural language. In our experience, writing SQL programs for this database is challenging since its schema is large and poorly documented. We aim to answer three key questions:
\begin{itemize}
\item Is it easy to write sketches in our language?
\item Is it easy to respond to questions?
\item How many iterations does our algorithm require?
\end{itemize}
We address the first two questions using a small-scale user studies where users write sketches or answer questions. We focus on evaluating feasibility of using our tool; additional experience would be needed for them to use it effectively.

\begin{figure}
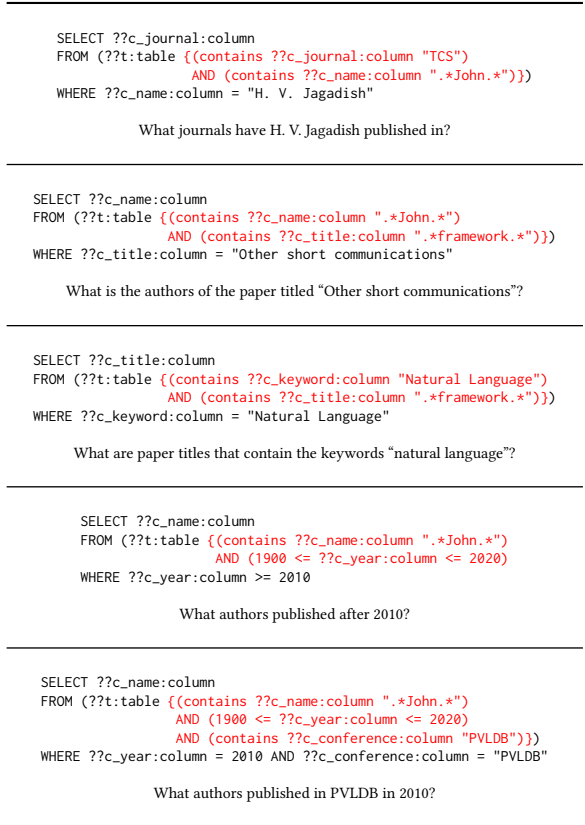

\centering\tiny
\begin{tabular}{c}
\toprule\\
$\begin{array}{l}
\verb+SELECT ??c_journal:column+ \\
\verb+FROM (??t:table +{\color{red}\texttt{\{(contains ??c\_journal:column "TCS")}} \\
\verb+                 +{\color{red}\texttt{AND (contains ??c\_name:column ".*John.*")\}}}\verb+)+ \\
\verb+WHERE ??c_name:column = "H. V. Jagadish"+ 
\end{array}$ \\\\
What journals have H. V. Jagadish published in? \\\\
\midrule\\
$\begin{array}{l}
\verb+SELECT ??c_name:column+ \\
\verb+FROM (??t:table +{\color{red}\texttt{\{(contains ??c\_name:column ".*John.*")}} \\
\verb+                 +{\color{red}\texttt{AND (contains ??c\_title:column ".*framework.*")\}}}\verb+)+ \\
\verb+WHERE ??c_title:column = "Other short communications"+ 
\end{array}$ \\\\
What is the authors of the paper titled ``Other short communications''? \\\\
\midrule\\
$\begin{array}{l}
\verb+SELECT ??c_title:column+ \\
\verb+FROM (??t:table +{\color{red}\texttt{\{(contains ??c\_keyword:column "Natural Language")}} \\
\verb+                 +{\color{red}\texttt{AND (contains ??c\_title:column ".*framework.*")\}}}\verb+)+ \\
\verb+WHERE ??c_keyword:column = "Natural Language"+\\
\end{array}$ \\\\
What are paper titles that contain the keywords ``natural language''? \\\\
\midrule\\
$\begin{array}{l}
\verb+SELECT ??c_name:column+ \\
\verb+FROM (??t:table +{\color{red}\texttt{\{(contains ??c\_name:column ".*John.*")}} \\
\verb+                 +{\color{red}\texttt{AND (1900 <= ??c\_year:column <= 2020)}} \\
\verb+WHERE ??c_year:column >= 2010+
\end{array}$ \\\\
What authors published after 2010? \\\\
\midrule\\
$\begin{array}{l}
\verb+SELECT ??c_name:column+ \\
\verb+FROM (??t:table +{\color{red}\texttt{\{(contains ??c\_name:column ".*John.*")}} \\
\verb+                 +{\color{red}\texttt{AND (1900 <= ??c\_year:column <= 2020)}} \\
\verb+                 +{\color{red}\texttt{AND (contains ??c\_conference:column "PVLDB")\}}}\verb+)+ \\
\verb+WHERE ??c_year:column = 2010 AND ??c_conference:column = "PVLDB"+
\end{array}$ \\\\
What authors published in PVLDB in 2010? \\\\
\bottomrule
\end{tabular}
\caption{Examples of sketches in our first user study. The soft constraints in each sketch are shown in red. The names of the column holes indicate the name of the column used to fill it, but these are ignored by our algorithm and can be changed without affected the semantics of the sketch.}
\label{fig:exsketch}
\end{figure}

\begin{figure}
\centering\tiny
\begin{tabular}{lc}
\toprule\\
$\begin{array}{l}
\verb+??c_name:column+
\end{array}
\xRightarrow{*}
\begin{array}{l}
\verb+aname+
\end{array}
$ & \true \\\\
\midrule\\
$\begin{array}{l}
\verb+??t:table+
\end{array}
\xRightarrow{*}
\begin{array}{l}
\verb+conference+ \\
\verb+INNER-JOIN ??t_new0:table+ \\
\verb+ON ??c_new0:column = ??c_new1:column+
\end{array}
$ & \true \\\\
\midrule\\
$\begin{array}{l}
\verb+??t_new0:table+
\end{array}
\xRightarrow{*}
\begin{array}{l}
\verb+publications+ \\
\verb+INNER-JOIN ??t_new1:table+ \\
\verb+ON ??c_new2:column = ??c_new3:column+
\end{array}
$ & \multirow{5}{*}{\true} \\
$\begin{array}{l}
\verb+??c_new0:column+
\end{array}
\xRightarrow{*}
\begin{array}{l}
\verb+cid+
\end{array}
$ & \\
$\begin{array}{l}
\verb+??c_new1:column+
\end{array}
\xRightarrow{*}
\begin{array}{l}
\verb+cid+
\end{array}
$ & \\\\
\midrule\\
$\begin{array}{l}
\verb+??t_new1:table+
\end{array}
\xRightarrow{*}
\begin{array}{l}
\verb+writes+ \\
\verb+INNER-JOIN ??t_new2:table+ \\
\verb+ON ??c_new4:column = ??c_new5:column+
\end{array}
$ & \multirow{5}{*}{\true} \\
$\begin{array}{l}
\verb+??c_new2:column+
\end{array}
\xRightarrow{*}
\begin{array}{l}
\verb+pid+
\end{array}
$ & \\
$\begin{array}{l}
\verb+??c_new3:column+
\end{array}
\xRightarrow{*}
\begin{array}{l}
\verb+pid+
\end{array}
$ & \\\\
\midrule\\
$\begin{array}{l}
\verb+??t_new2:table+
\end{array}
\xRightarrow{*}
\begin{array}{l}
\verb+domain_author+
\end{array}
$ & \multirow{5}{*}{\false} \\
$\begin{array}{l}
\verb+??c_new2:column+
\end{array}
\xRightarrow{*}
\begin{array}{l}
\verb+pid+
\end{array}
$ & \\
$\begin{array}{l}
\verb+??c_new3:column+
\end{array}
\xRightarrow{*}
\begin{array}{l}
\verb+pid+
\end{array}
$ & \\\\
\midrule\\
\multicolumn{2}{c}{$...$} \\\\
\bottomrule
\end{tabular}
\caption{Sequence of questions for the first sketch in Figure~\ref{fig:exsketch}. We show the question (left) and our response (right).}
\label{fig:exquery}
\end{figure}

\begin{figure*}
\centering
\includegraphics[width=0.3\textwidth]{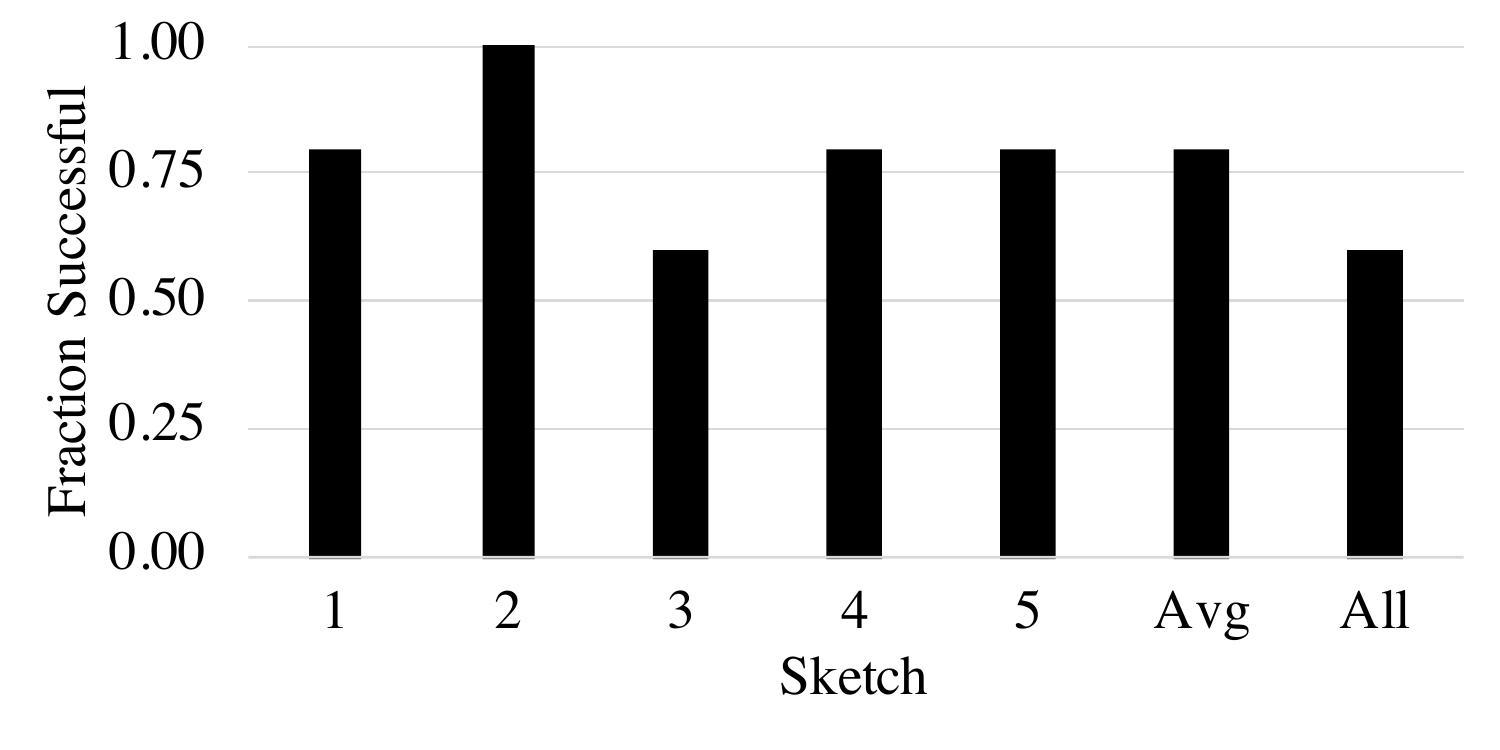}
\includegraphics[width=0.3\textwidth]{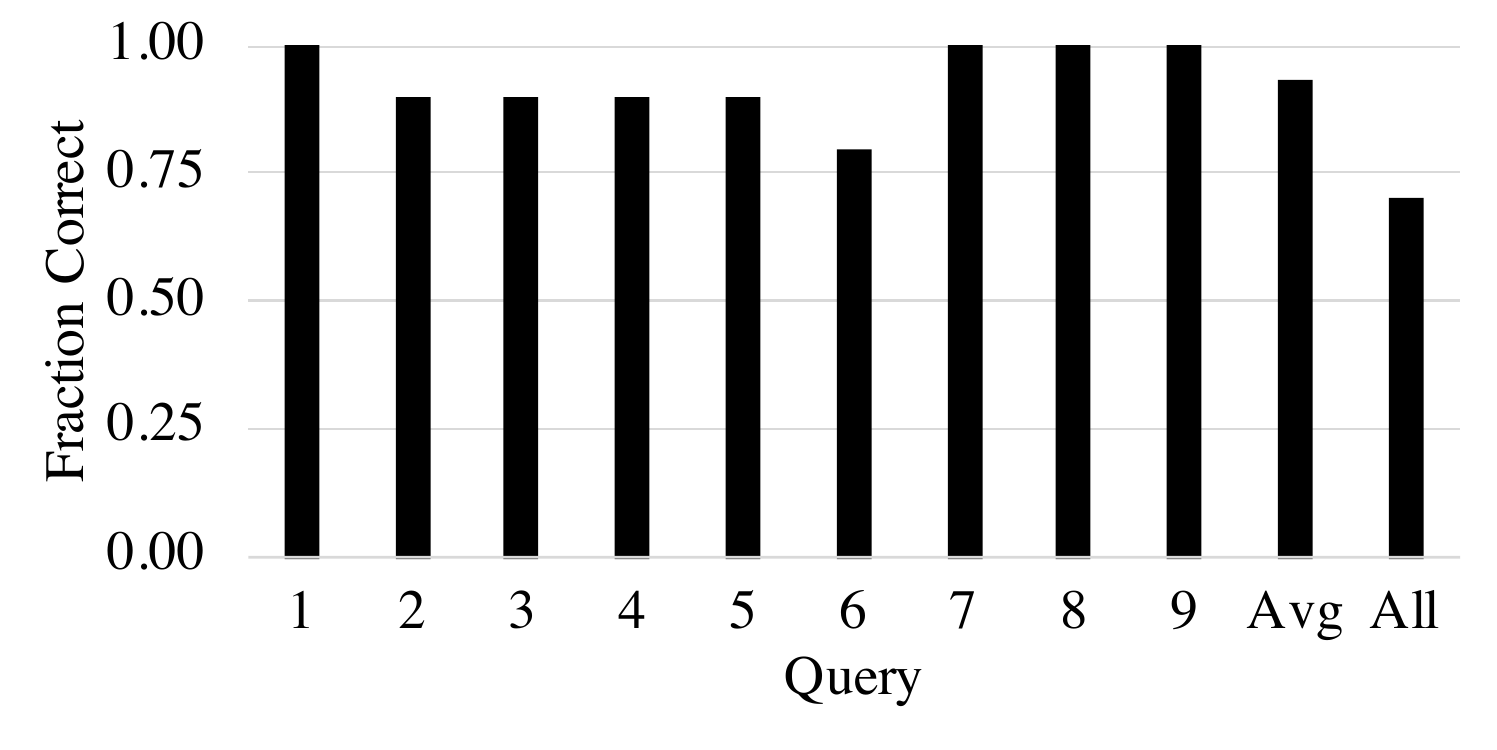}
\includegraphics[width=0.3\textwidth]{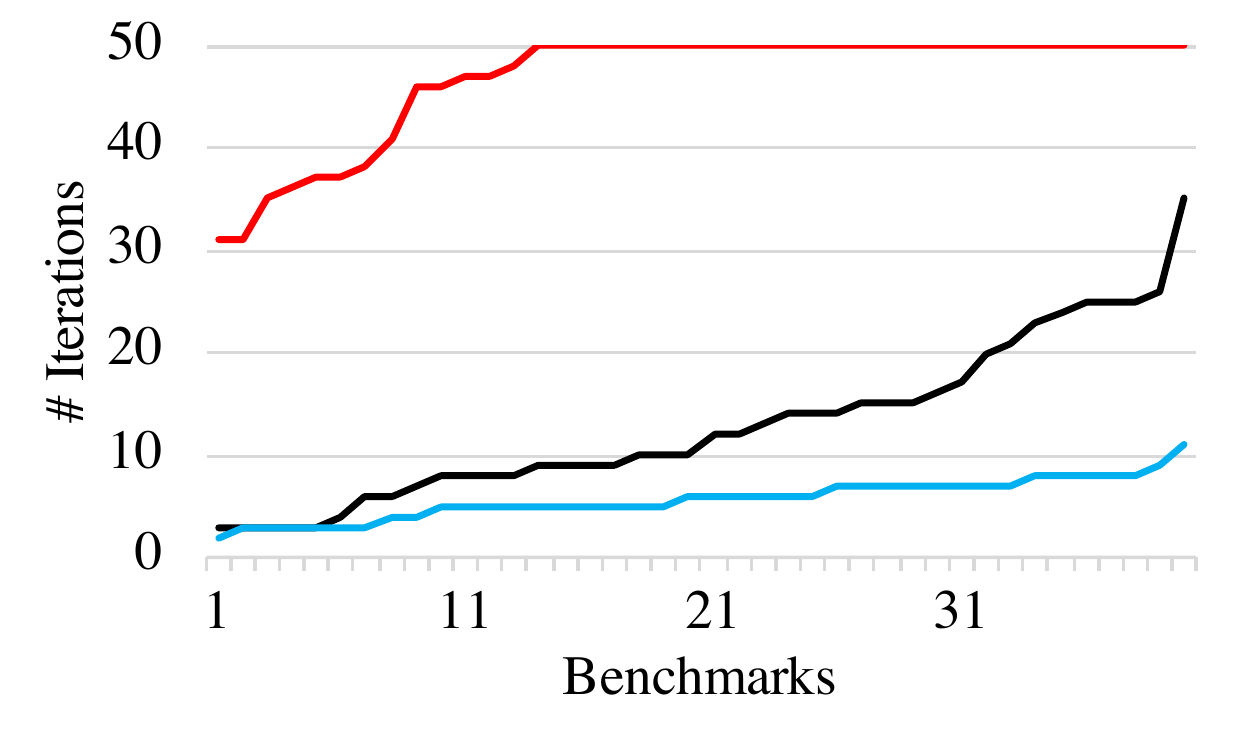}
\caption{Left: Fraction of user-written sketches that are successful at solving the task; ``Avg'' is the average across the five sketches, and ``All'' is the fraction of users that provided all successful sketches. Middle: Fraction of users that responded correctly to questions; ``Avg'' is the average across the nine questions, and ``All'' is the fraction of users that responded correctly to all questions. Right: Number of iterations used by our algorithm (black), a baseline that omits soft constraints (red), and an oracle that only asks correct questions (blue). The benchmarks are sorted by number of iterations (independently for each of the three curves). We time out benchmarks at 50 iterations or 1 hour.}
\label{fig:exiters}
\end{figure*}

\paragraph{Writing sketches.}

We performed a user study to evaluate whether users can write effective sketches; in particular, our goal is to evaluate whether they can write specifications for an SQL program without seeing the contents of the database.

First, we selected natural language descriptions of five SQL programs, choosing each one to include a single new column compared to the previous ones for which the user needed to provide a soft constraint. For each of the five SQL programs, we wrote a corresponding ``ground truth'' sketch; for all of these, our tool terminated in at most ten iterations. Figure~\ref{fig:exsketch} shows each ground truth sketch along with a natural language description of the corresponding SQL program.

Each ground truth sketch includes at least one soft constraint for each column that appears in the sketch---i.e., each column in the project operation or in the logical formula of the select operation. Intuitively, this strategy ensures that each column hole is at least somewhat constrained, enabling our algorithm to quickly identify which columns it should include in the flat table along with the appropriate sequence of inner-join operations. Since we expect each of these columns to occur in the flat table constructed by the table hole; thus, we added these soft constraints on the table hole. For example, for the column of author names, we expected some author to be named \texttt{John}, so we added a soft constraint \texttt{(contains ?? ".*John.*)} to the table hole in that sketch.

Then, we recruited five users who were familiar with writing SQL programs. We showed them the natural language description for each of the five chosen SQL programs and asked them to write a sketch that captures its semantics. To train them to use our system, we asked them to read through and understand a detailed example of a sketch for an SQL program. We used an SQL program for a database from another domain (in particular, an SQL program targeting a healthcare database) to ensure that there was little or no overlap in terms of the nature of the specific strategies used to write specifications, though it conveyed qualities that make a specification effective. In addition, we gave them the above guideline for writing effective sketches---i.e., that it should include at least one soft constraint for each column.

For this part of our evaluation, we did not provide users with interactive feedback for their sketch. One resulting issue was that since the users did not have access to a syntax checker, they made syntactic mistakes; we manually fixed these mistakes as long as doing so did not affect the intended semantics of the sketch. Then, we ran each of these sketches using \toolname, automatically responding to its questions based on whether they matched the ground truth sketch written by ourselves. We count a sketch as successful if it produces the correct SQL program in within two iterations of the number of iterations required when using the ground truth sketch.

In Figure~\ref{fig:exiters} (left), we show the success rate of each sketch; the average success rate is 80\%. Also, for three of the five users, all five sketches were successful. These results show that users can write successful sketches without significant training. Investigating the failure cases, we found they were all due to a failure to follow the guideline that a soft constraint should be included for each column used in the sketch. Since users could not interact with \toolname in this part of our study, we believe they were unable to internalize our guideline. In particular, one of our three users followed this strategy and wrote a successful sketch each of the five queries.

\paragraph{Responding to questions.}

Next, we evaluate whether users can respond to questions made by \toolname during sketch refinement. This study was significantly shorter, so we had a higher response rate of ten users. We show each user a description of an SQL program from the healthcare domain, along with an example of a question that should be accepted and one that should be rejected. Then, using the last ground truth sketch in Figure~\ref{fig:exsketch}, we asked them respond to the sequence of questions asked by \toolname. If all questions are answered correctly, there are nine total questions that the user must respond; the first five questions are shown in Figure~\ref{fig:exquery}. For seven of these, the correct response is ``accept'' (i.e., the proposed refinement was correct), and for two the response is ``reject'' (i.e., the proposed refinement was incorrect).

To help the user make their decision, we showed them the current sketch, the candidate refinement, and the first few rows of the relevant tables. For a column used to fill a column hole, we show them the table containing the candidate refinement. For an table hole, if the candidate refinement proposes to fill it with an inner-join operation, we show them both tables involved in that inner-join; otherwise, if it proposes filling it with a single table, we show them that table---e.g., for the second question, we show the table \texttt{conference}:
\begin{center}
\scriptsize
\begin{tabular}{rlll}
\toprule
\multicolumn{1}{c}{{\bf cid}} & \multicolumn{1}{c}{{\bf cname}} & \multicolumn{1}{c}{{\bf full\_name}} & \multicolumn{1}{c}{{\bf homepage}} \\
\midrule
232 & DAC & Design Automation Conference & \multicolumn{1}{c}{...} \\
134 & ATC & Autonomic and Trusted Computing & \multicolumn{1}{c}{...} \\
\multicolumn{4}{c}{...} \\
\bottomrule
\end{tabular}
\end{center}
Based on this information, it is apparent that this column is the correct one, so this question should be accepted. 

For each of the nine questions, we evaluate how many users can answer them correctly, along with how many users answer \emph{all} nine correctly; the latter measures the number of users that would obtain the correct sketch. Results are shown in Figure~\ref{fig:exiters} (middle). Seven of the ten users correctly answered all questions; thus, our results suggest that users can easily respond to questions issued by \toolname.

\paragraph{Iterations required.}

We study the number of iterations required for our algorithm to converge to the true program $\complete{P}^*$. We run our algorithm using each of the sketches we wrote as input. We compare to a baseline where our algorithm is run with the soft constraints omitted. This baseline helps capture the size of the search space---in particular, it captures how our algorithm can use the soft constraints to cut down the search space. We also compare to an oracle that only asks questions for which the user responds \true---i.e., a measure of the minimum amount of work that the user must do.

We show results in Figure~\ref{fig:exiters}. Our algorithm terminates in fewer than 30 iterations in all but one case, and in fewer than 20 iterations in more than 75\% of cases. The one case that took more than 30 iterations includes a sequence of five tables inner-joined together. In contrast, the baseline takes a large number of iterations---it times out on 27 of the 40 benchmarks (we time out after 50 iterations or 1 hour). Also, for the most part, our algorithm is only a factor of two worse than the oracle, and furthermore matches the oracle for easy benchmarks (i.e., those at the left-hand side of the plot). In contrast, the baseline is an order of magnitude worse even for easy benchmarks. Thus, our algorithm substantially cuts down the search space compared to the baseline.

\paragraph{Limitations.}

We briefly discuss a few of the limitations in our system. First, \toolname only implements a fragment of SQL---in particular, it omits aggregation operations. Extending our approach to work with these operations is straightforward; once implemented, the user can include these additional operations in the sketch (similarly to projection and selection operations), and our algorithm would fill in the holes with columns, tables, and inner-join operations.

Another limitation is the limited number of constraints that we provide. We have deliberately omitted soft constraints on words in the column names since these names may be misleading in practice. Nevertheless, it is easy to extend our system to include additional soft constraints.

Finally, our algorithm restricts to holes that are either sequences of inner-join operations (i.e., nonterminals $I$) or columns (i.e., nonterminals $C$). One particular place where a user might want to include a hole is for the constants that appear in the logical formula $\psi$ in a select operation $\sigma_{\psi}(t)$. For example, in $P_{\text{author}}$, there is a select operation $\sigma_{C:\texttt{c\_year}=1948}(...)$. In this example, the user might not be sure how the current year is expressed, in which case they would be unable to write this sketch. However, this restriction is easy to address---the user can simply use a temporary logical formula $\psi=\true$ in the sketch to synthesize the program. Then, they can inspect the column in the synthesized program to determine the appropriate way to write $\psi$.

\section{Related Work}

\paragraph{Program synthesis.}

There has been recent interest in program synthesis. We can divide the literature along two dimensions: (i) the kind of user specifications that are used, and (ii) the search strategy. In terms of user specifications, there has been work on logical specifications~\cite{alur2013syntax,polikarpova2016program}, sketches (i.e., a logical specification along with a sketch specifying the high-level structure of the code)~\cite{solar2005programming,solar2006combinatorial,solar2009sketching}, input-output examples~\cite{gulwani2011automating,feser2015synthesizing,osera2015type,polozov2015flashmeta,feng2017component,wang2017synthesizing}, and natural language~\cite{yaghmazadeh2017sqlizer}. 

Input-output examples and natural language specifications are especially prone to ambiguity---i.e., there are multiple programs with different semantics that satisfy the specification. In these cases, interaction has been used to resolve ambiguity~\cite{li2014constructing,le2017interactive,bastani2017synthesizing,wang2017interactive,drachsler2017synthesis,pu2018selecting,bastani2018active,ji2020interactive}. For the most part, these approaches have largely focused on obtaining additional input-output examples; as a consequence, they are typically heuristic and are unable to ensure correctness. One approach uses \emph{abstract input-output examples}, which represent a potentially infinite set of concrete input-output examples; however, giving input-output examples is our setting is not practical, since we assume the user does not know the database schema. The most closely related work is~\cite{li2014constructing}, which interacts with the user to resolve ambiguity in user-provided natural language description. In contrast, our approach allows the user to provide more powerful specifications compared to natural language, and also has correctness guarantees.

In terms of search strategy, there has been work on deductive search~\cite{manna1986deductive}, constraint-based search~\cite{solar2005programming,solar2006combinatorial,solar2009sketching}, search using version space algebras~\cite{gulwani2011automating,polozov2015flashmeta}, and (guided) enumerative search~\cite{alur2013syntax,feser2015synthesizing}. We build on enumerative search in the context of syntax-guided synthesis~\cite{alur2013syntax}, where the search is guided by a context-free grammar encoding the semantics of a domain-specific language. In contrast to existing approaches, however, our algorithm needs to sample programs rather than find a single one that satisfies the specification. On the one hand, our problem is more challenging since we need to generate many complete programs; on the other hand, it is easier since we do not need to find the true program on early iterations of our algorithm.

Next, there has been work on using AI to guide program synthesis~\cite{balog2016deepcoder,feng2018program,kalyan2018neural,si2018learning2,lee2018accelerating}, including the use of stochastic search~\cite{schkufza2013stochastic,schkufza2014stochastic}. Our usage of stochastic search is different---we use it to generate samples from a distribution to use to select questions, rather than to optimize an objective. Finally, there has been work on synthesizing database queries from natural language~\cite{li2014constructing,yaghmazadeh2017sqlizer} and from examples~\cite{feng2017component,wang2017synthesizing,martins12trinity,yaghmazadeh2018automated}.

\paragraph{Refinement programming.}

There has been work on a refinement approach to writing programs~\cite{dijkstra1968constructive,back1978correctness,morgan1994programming,klein2009sel4}. The goal is typically to develop programs that are correct by construction, where the correctness proofs are written alongside the program. In this approach, abstract components (which correspond to holes in this setting) become progressively more concrete. However, these approaches are largely manual and do not leverage interaction---i.e., the programmer has to manually both decide which abstract components to refine and implement the refinement. In contrast, our algorithm actively proposes candidate refinements, and the user only needs to accept or reject these candidates.

\section{Conclusion}

We have proposed a novel approach to interactively synthesizing database queries. Our approach is based on interactively refining a user-provided sketch with soft constraints on the expected semantics. By leveraging rich user feedback, our approach is able to provide strong correctness guarantees. We show how our algorithm can be used to synthesize queries for a database of academic publications. Future work includes implementing additional database operations and soft constraints, improving the sampling algorithm, learning a prior over programs to help guide the search, designing a user interface to facilitate interactions, and applying our approach to other domains.

\bibliography{main}


\begin{thebibliography}{36}


\ifx \showCODEN    \undefined \def \showCODEN     #1{\unskip}     \fi
\ifx \showDOI      \undefined \def \showDOI       #1{#1}\fi
\ifx \showISBNx    \undefined \def \showISBNx     #1{\unskip}     \fi
\ifx \showISBNxiii \undefined \def \showISBNxiii  #1{\unskip}     \fi
\ifx \showISSN     \undefined \def \showISSN      #1{\unskip}     \fi
\ifx \showLCCN     \undefined \def \showLCCN      #1{\unskip}     \fi
\ifx \shownote     \undefined \def \shownote      #1{#1}          \fi
\ifx \showarticletitle \undefined \def \showarticletitle #1{#1}   \fi
\ifx \showURL      \undefined \def \showURL       {\relax}        \fi
\providecommand\bibfield[2]{#2}
\providecommand\bibinfo[2]{#2}
\providecommand\natexlab[1]{#1}
\providecommand\showeprint[2][]{arXiv:#2}

\bibitem[\protect\citeauthoryear{Alur, Bodik, Juniwal, Martin, Raghothaman,
  Seshia, Singh, Sola\~r Lezama, Torlak, and Udupa}{Alur et~al\mbox{.}}{2013}]%
        {alur2013syntax}
\bibfield{author}{\bibinfo{person}{Rajeev Alur}, \bibinfo{person}{Rastislav
  Bodik}, \bibinfo{person}{Garvit Juniwal}, \bibinfo{person}{Milo~MK Martin},
  \bibinfo{person}{Mukund Raghothaman}, \bibinfo{person}{Sanjit~A Seshia},
  \bibinfo{person}{Rishabh Singh}, \bibinfo{person}{Armando Sola\~r Lezama},
  \bibinfo{person}{Emina Torlak}, {and} \bibinfo{person}{Abhishek Udupa}.}
  \bibinfo{year}{2013}\natexlab{}.
\newblock \showarticletitle{Syntax-guided synthesis}. IEEE,
  \bibinfo{pages}{1--8}.
\newblock


\bibitem[\protect\citeauthoryear{Back}{Back}{1978}]%
        {back1978correctness}
\bibfield{author}{\bibinfo{person}{Ralph~Johan Back}.}
  \bibinfo{year}{1978}\natexlab{}.
\newblock \bibinfo{booktitle}{\emph{On the correctness of refinement steps in
  program development}}.
\newblock \bibinfo{publisher}{Department of Computer Science, University of
  Helsinki Helsinki, Finland}.
\newblock


\bibitem[\protect\citeauthoryear{Balog, Gaunt, Brockschmidt, Nowozin, and
  Tarlow}{Balog et~al\mbox{.}}{2016}]%
        {balog2016deepcoder}
\bibfield{author}{\bibinfo{person}{Matej Balog}, \bibinfo{person}{Alexander~L
  Gaunt}, \bibinfo{person}{Marc Brockschmidt}, \bibinfo{person}{Sebastian
  Nowozin}, {and} \bibinfo{person}{Daniel Tarlow}.}
  \bibinfo{year}{2016}\natexlab{}.
\newblock \showarticletitle{Deepcoder: Learning to write programs}. In
  \bibinfo{booktitle}{\emph{ICLR}}.
\newblock


\bibitem[\protect\citeauthoryear{Bastani, Sharma, Aiken, and Liang}{Bastani
  et~al\mbox{.}}{2017}]%
        {bastani2017synthesizing}
\bibfield{author}{\bibinfo{person}{Osbert Bastani}, \bibinfo{person}{Rahul
  Sharma}, \bibinfo{person}{Alex Aiken}, {and} \bibinfo{person}{Percy Liang}.}
  \bibinfo{year}{2017}\natexlab{}.
\newblock \showarticletitle{Synthesizing program input grammars}. In
  \bibinfo{booktitle}{\emph{PLDI}}, Vol.~\bibinfo{volume}{52}. ACM,
  \bibinfo{pages}{95--110}.
\newblock


\bibitem[\protect\citeauthoryear{Bastani, Sharma, Aiken, and Liang}{Bastani
  et~al\mbox{.}}{2018}]%
        {bastani2018active}
\bibfield{author}{\bibinfo{person}{Osbert Bastani}, \bibinfo{person}{Rahul
  Sharma}, \bibinfo{person}{Alex Aiken}, {and} \bibinfo{person}{Percy Liang}.}
  \bibinfo{year}{2018}\natexlab{}.
\newblock \showarticletitle{Active learning of points-to specifications}. In
  \bibinfo{booktitle}{\emph{Proceedings of the 39th ACM SIGPLAN Conference on
  Programming Language Design and Implementation}}. ACM,
  \bibinfo{pages}{678--692}.
\newblock


\bibitem[\protect\citeauthoryear{Chib and Greenberg}{Chib and
  Greenberg}{1995}]%
        {chib1995understanding}
\bibfield{author}{\bibinfo{person}{Siddhartha Chib} {and}
  \bibinfo{person}{Edward Greenberg}.} \bibinfo{year}{1995}\natexlab{}.
\newblock \showarticletitle{Understanding the metropolis-hastings algorithm}.
\newblock \bibinfo{journal}{\emph{The american statistician}}
  \bibinfo{volume}{49}, \bibinfo{number}{4} (\bibinfo{year}{1995}),
  \bibinfo{pages}{327--335}.
\newblock


\bibitem[\protect\citeauthoryear{Dasgupta}{Dasgupta}{2005}]%
        {dasgupta2005analysis}
\bibfield{author}{\bibinfo{person}{Sanjoy Dasgupta}.}
  \bibinfo{year}{2005}\natexlab{}.
\newblock \showarticletitle{Analysis of a greedy active learning strategy}. In
  \bibinfo{booktitle}{\emph{Advances in neural information processing
  systems}}. \bibinfo{pages}{337--344}.
\newblock


\bibitem[\protect\citeauthoryear{Dijkstra}{Dijkstra}{1968}]%
        {dijkstra1968constructive}
\bibfield{author}{\bibinfo{person}{Edsger~W Dijkstra}.}
  \bibinfo{year}{1968}\natexlab{}.
\newblock \showarticletitle{A constructive approach to the problem of program
  correctness}.
\newblock \bibinfo{journal}{\emph{BIT Numerical Mathematics}}
  \bibinfo{volume}{8}, \bibinfo{number}{3} (\bibinfo{year}{1968}),
  \bibinfo{pages}{174--186}.
\newblock


\bibitem[\protect\citeauthoryear{Drachsler-Cohen, Shoham, and
  Yahav}{Drachsler-Cohen et~al\mbox{.}}{2017}]%
        {drachsler2017synthesis}
\bibfield{author}{\bibinfo{person}{Dana Drachsler-Cohen},
  \bibinfo{person}{Sharon Shoham}, {and} \bibinfo{person}{Eran Yahav}.}
  \bibinfo{year}{2017}\natexlab{}.
\newblock \showarticletitle{Synthesis with abstract examples}. In
  \bibinfo{booktitle}{\emph{International Conference on Computer Aided
  Verification}}. \bibinfo{pages}{254--278}.
\newblock


\bibitem[\protect\citeauthoryear{Feng, Martins, Bastani, and Dillig}{Feng
  et~al\mbox{.}}{2018}]%
        {feng2018program}
\bibfield{author}{\bibinfo{person}{Yu Feng}, \bibinfo{person}{Ruben Martins},
  \bibinfo{person}{Osbert Bastani}, {and} \bibinfo{person}{Isil Dillig}.}
  \bibinfo{year}{2018}\natexlab{}.
\newblock \showarticletitle{Program synthesis using conflict-driven learning}.
  In \bibinfo{booktitle}{\emph{Proceedings of the 39th ACM SIGPLAN Conference
  on Programming Language Design and Implementation}}. ACM,
  \bibinfo{pages}{420--435}.
\newblock


\bibitem[\protect\citeauthoryear{Feng, Martins, Van~Geffen, Dillig, and
  Chaudhuri}{Feng et~al\mbox{.}}{2017}]%
        {feng2017component}
\bibfield{author}{\bibinfo{person}{Yu Feng}, \bibinfo{person}{Ruben Martins},
  \bibinfo{person}{Jacob Van~Geffen}, \bibinfo{person}{Isil Dillig}, {and}
  \bibinfo{person}{Swarat Chaudhuri}.} \bibinfo{year}{2017}\natexlab{}.
\newblock \showarticletitle{Component-based synthesis of table consolidation
  and transformation tasks from examples}. In \bibinfo{booktitle}{\emph{PLDI}},
  Vol.~\bibinfo{volume}{52}. \bibinfo{publisher}{ACM},
  \bibinfo{pages}{422--436}.
\newblock


\bibitem[\protect\citeauthoryear{Feser, Chaudhuri, and Dillig}{Feser
  et~al\mbox{.}}{2015}]%
        {feser2015synthesizing}
\bibfield{author}{\bibinfo{person}{John~K. Feser}, \bibinfo{person}{Swarat
  Chaudhuri}, {and} \bibinfo{person}{Isil Dillig}.}
  \bibinfo{year}{2015}\natexlab{}.
\newblock \showarticletitle{{Synthesizing Data Structure Transformations from
  Input-output Examples}}. \bibinfo{publisher}{ACM}, \bibinfo{pages}{229--239}.
\newblock


\bibitem[\protect\citeauthoryear{Gulwani}{Gulwani}{2011}]%
        {gulwani2011automating}
\bibfield{author}{\bibinfo{person}{Sumit Gulwani}.}
  \bibinfo{year}{2011}\natexlab{}.
\newblock \showarticletitle{Automating string processing in spreadsheets using
  input-output examples}. \bibinfo{publisher}{ACM}, \bibinfo{pages}{317--330}.
\newblock


\bibitem[\protect\citeauthoryear{Ji, Liang, Xiong, Zhang, and Hu}{Ji
  et~al\mbox{.}}{2020}]%
        {ji2020interactive}
\bibfield{author}{\bibinfo{person}{Ruyi Ji}, \bibinfo{person}{Jingjing Liang},
  \bibinfo{person}{Yingfei Xiong}, \bibinfo{person}{Lu Zhang}, {and}
  \bibinfo{person}{Zhenjiang Hu}.} \bibinfo{year}{2020}\natexlab{}.
\newblock \showarticletitle{Question selection for interactive program
  synthesis}. In \bibinfo{booktitle}{\emph{Proceedings of the 41st {ACM}
  {SIGPLAN} International Conference on Programming Language Design and
  Implementation}}. \bibinfo{publisher}{{ACM}}, \bibinfo{pages}{1143--1158}.
\newblock


\bibitem[\protect\citeauthoryear{Kalyan, Mohta, Polozov, Batra, Jain, and
  Gulwani}{Kalyan et~al\mbox{.}}{2018}]%
        {kalyan2018neural}
\bibfield{author}{\bibinfo{person}{Ashwin Kalyan}, \bibinfo{person}{Abhishek
  Mohta}, \bibinfo{person}{Oleksandr Polozov}, \bibinfo{person}{Dhruv Batra},
  \bibinfo{person}{Prateek Jain}, {and} \bibinfo{person}{Sumit Gulwani}.}
  \bibinfo{year}{2018}\natexlab{}.
\newblock \showarticletitle{Neural-Guided Deductive Search for Real-Time
  Program Synthesis from Examples}. In \bibinfo{booktitle}{\emph{ICLR}}.
\newblock


\bibitem[\protect\citeauthoryear{Klein, Elphinstone, Heiser, Andronick, Cock,
  Derrin, Elkaduwe, Engelhardt, Kolanski, Norrish, et~al\mbox{.}}{Klein
  et~al\mbox{.}}{2009}]%
        {klein2009sel4}
\bibfield{author}{\bibinfo{person}{Gerwin Klein}, \bibinfo{person}{Kevin
  Elphinstone}, \bibinfo{person}{Gernot Heiser}, \bibinfo{person}{June
  Andronick}, \bibinfo{person}{David Cock}, \bibinfo{person}{Philip Derrin},
  \bibinfo{person}{Dhammika Elkaduwe}, \bibinfo{person}{Kai Engelhardt},
  \bibinfo{person}{Rafal Kolanski}, \bibinfo{person}{Michael Norrish},
  {et~al\mbox{.}}} \bibinfo{year}{2009}\natexlab{}.
\newblock \showarticletitle{seL4: Formal verification of an OS kernel}. In
  \bibinfo{booktitle}{\emph{Proceedings of the ACM SIGOPS 22nd symposium on
  Operating systems principles}}. ACM, \bibinfo{pages}{207--220}.
\newblock


\bibitem[\protect\citeauthoryear{Le, Perelman, Polozov, Raza, Udupa, and
  Gulwani}{Le et~al\mbox{.}}{2017}]%
        {le2017interactive}
\bibfield{author}{\bibinfo{person}{Vu Le}, \bibinfo{person}{Daniel Perelman},
  \bibinfo{person}{Oleksandr Polozov}, \bibinfo{person}{Mohammad Raza},
  \bibinfo{person}{Abhishek Udupa}, {and} \bibinfo{person}{Sumit Gulwani}.}
  \bibinfo{year}{2017}\natexlab{}.
\newblock \showarticletitle{Interactive Program Synthesis}.
\newblock \bibinfo{journal}{\emph{arXiv preprint arXiv:1703.03539}}
  (\bibinfo{year}{2017}).
\newblock


\bibitem[\protect\citeauthoryear{Lee, Heo, Alur, and Naik}{Lee
  et~al\mbox{.}}{2018}]%
        {lee2018accelerating}
\bibfield{author}{\bibinfo{person}{Woosuk Lee}, \bibinfo{person}{Kihong Heo},
  \bibinfo{person}{Rajeev Alur}, {and} \bibinfo{person}{Mayur Naik}.}
  \bibinfo{year}{2018}\natexlab{}.
\newblock \showarticletitle{Accelerating search-based program synthesis using
  learned probabilistic models}. In \bibinfo{booktitle}{\emph{Proceedings of
  the 39th ACM SIGPLAN Conference on Programming Language Design and
  Implementation}}. ACM, \bibinfo{pages}{436--449}.
\newblock


\bibitem[\protect\citeauthoryear{Li and Jagadish}{Li and Jagadish}{2014}]%
        {li2014constructing}
\bibfield{author}{\bibinfo{person}{Fei Li} {and} \bibinfo{person}{HV
  Jagadish}.} \bibinfo{year}{2014}\natexlab{}.
\newblock \showarticletitle{Constructing an interactive natural language
  interface for relational databases}.
\newblock \bibinfo{journal}{\emph{Proceedings of the VLDB Endowment}}
  \bibinfo{volume}{8}, \bibinfo{number}{1} (\bibinfo{year}{2014}),
  \bibinfo{pages}{73--84}.
\newblock


\bibitem[\protect\citeauthoryear{Manna and Waldinger}{Manna and
  Waldinger}{1986}]%
        {manna1986deductive}
\bibfield{author}{\bibinfo{person}{Zohar Manna} {and} \bibinfo{person}{Richard
  Waldinger}.} \bibinfo{year}{1986}\natexlab{}.
\newblock \showarticletitle{A deductive approach to program synthesis}.
\newblock In \bibinfo{booktitle}{\emph{Readings in artificial intelligence and
  software engineering}}. \bibinfo{publisher}{Elsevier},
  \bibinfo{pages}{3--34}.
\newblock


\bibitem[\protect\citeauthoryear{Martins, Chen, Chen, Feng, and Dillig}{Martins
  et~al\mbox{.}}{[n. d.]}]%
        {martins12trinity}
\bibfield{author}{\bibinfo{person}{Ruben Martins}, \bibinfo{person}{Jia Chen},
  \bibinfo{person}{Yanju Chen}, \bibinfo{person}{Yu Feng}, {and}
  \bibinfo{person}{Isil Dillig}.} \bibinfo{year}{[n. d.]}\natexlab{}.
\newblock \showarticletitle{Trinity: An Extensible Synthesis Framework for Data
  Science}.
\newblock \bibinfo{journal}{\emph{Proceedings of the VLDB Endowment}}
  \bibinfo{volume}{12}, \bibinfo{number}{12} (\bibinfo{year}{[n. d.]}).
\newblock


\bibitem[\protect\citeauthoryear{Morgan}{Morgan}{1994}]%
        {morgan1994programming}
\bibfield{author}{\bibinfo{person}{Carroll Morgan}.}
  \bibinfo{year}{1994}\natexlab{}.
\newblock \bibinfo{booktitle}{\emph{Programming from specifications}}.
\newblock \bibinfo{publisher}{Prentice Hall,}.
\newblock


\bibitem[\protect\citeauthoryear{Osera and Zdancewic}{Osera and
  Zdancewic}{2015}]%
        {osera2015type}
\bibfield{author}{\bibinfo{person}{Peter-Michael Osera} {and}
  \bibinfo{person}{Steve Zdancewic}.} \bibinfo{year}{2015}\natexlab{}.
\newblock \showarticletitle{Type-and-example-directed program synthesis}. ACM,
  \bibinfo{pages}{619--630}.
\newblock


\bibitem[\protect\citeauthoryear{Polikarpova, Kuraj, and
  Solar-Lezama}{Polikarpova et~al\mbox{.}}{2016}]%
        {polikarpova2016program}
\bibfield{author}{\bibinfo{person}{Nadia Polikarpova}, \bibinfo{person}{Ivan
  Kuraj}, {and} \bibinfo{person}{Armando Solar-Lezama}.}
  \bibinfo{year}{2016}\natexlab{}.
\newblock \showarticletitle{Program synthesis from polymorphic refinement
  types}.
\newblock  (\bibinfo{year}{2016}), \bibinfo{pages}{522--538}.
\newblock


\bibitem[\protect\citeauthoryear{Polozov and Gulwani}{Polozov and
  Gulwani}{2015}]%
        {polozov2015flashmeta}
\bibfield{author}{\bibinfo{person}{Oleksandr Polozov} {and}
  \bibinfo{person}{Sumit Gulwani}.} \bibinfo{year}{2015}\natexlab{}.
\newblock \showarticletitle{FlashMeta: a framework for inductive program
  synthesis}. In \bibinfo{booktitle}{\emph{ACM SIGPLAN Notices}},
  Vol.~\bibinfo{volume}{50}. ACM, \bibinfo{pages}{107--126}.
\newblock


\bibitem[\protect\citeauthoryear{Pu, Miranda, Solar-Lezama, and Kaelbling}{Pu
  et~al\mbox{.}}{2018}]%
        {pu2018selecting}
\bibfield{author}{\bibinfo{person}{Yewen Pu}, \bibinfo{person}{Zachery
  Miranda}, \bibinfo{person}{Armando Solar-Lezama}, {and}
  \bibinfo{person}{Leslie~Pack Kaelbling}.} \bibinfo{year}{2018}\natexlab{}.
\newblock \showarticletitle{Selecting representative examples for program
  synthesis}. In \bibinfo{booktitle}{\emph{ICML}}.
\newblock


\bibitem[\protect\citeauthoryear{Schkufza, Sharma, and Aiken}{Schkufza
  et~al\mbox{.}}{2013}]%
        {schkufza2013stochastic}
\bibfield{author}{\bibinfo{person}{Eric Schkufza}, \bibinfo{person}{Rahul
  Sharma}, {and} \bibinfo{person}{Alex Aiken}.}
  \bibinfo{year}{2013}\natexlab{}.
\newblock \showarticletitle{Stochastic superoptimization}. In
  \bibinfo{booktitle}{\emph{ASPLOS}}, Vol.~\bibinfo{volume}{41}. ACM,
  \bibinfo{pages}{305--316}.
\newblock


\bibitem[\protect\citeauthoryear{Schkufza, Sharma, and Aiken}{Schkufza
  et~al\mbox{.}}{2014}]%
        {schkufza2014stochastic}
\bibfield{author}{\bibinfo{person}{Eric Schkufza}, \bibinfo{person}{Rahul
  Sharma}, {and} \bibinfo{person}{Alex Aiken}.}
  \bibinfo{year}{2014}\natexlab{}.
\newblock \showarticletitle{Stochastic optimization of floating-point programs
  with tunable precision}. In \bibinfo{booktitle}{\emph{PLDI}},
  Vol.~\bibinfo{volume}{49}. \bibinfo{publisher}{ACM}, \bibinfo{pages}{53--64}.
\newblock


\bibitem[\protect\citeauthoryear{Si, Yang, Dai, Naik, and Song}{Si
  et~al\mbox{.}}{2018}]%
        {si2018learning2}
\bibfield{author}{\bibinfo{person}{Xujie Si}, \bibinfo{person}{Yuan Yang},
  \bibinfo{person}{Hanjun Dai}, \bibinfo{person}{Mayur Naik}, {and}
  \bibinfo{person}{Le Song}.} \bibinfo{year}{2018}\natexlab{}.
\newblock \showarticletitle{Learning a Meta-Solver for Syntax-Guided Program
  Synthesis}. In \bibinfo{booktitle}{\emph{ICLR}}.
\newblock


\bibitem[\protect\citeauthoryear{Solar-Lezama}{Solar-Lezama}{2009}]%
        {solar2009sketching}
\bibfield{author}{\bibinfo{person}{Armando Solar-Lezama}.}
  \bibinfo{year}{2009}\natexlab{}.
\newblock \showarticletitle{The Sketching Approach to Program Synthesis.}. In
  \bibinfo{booktitle}{\emph{Proc. Asian Symposium on Programming Languages and
  Systems}}. Springer, \bibinfo{pages}{4--13}.
\newblock


\bibitem[\protect\citeauthoryear{Solar{-}Lezama, Rabbah, Bod{\'{\i}}k, and
  Ebcioglu}{Solar{-}Lezama et~al\mbox{.}}{2005}]%
        {solar2005programming}
\bibfield{author}{\bibinfo{person}{Armando Solar{-}Lezama},
  \bibinfo{person}{Rodric~M. Rabbah}, \bibinfo{person}{Rastislav Bod{\'{\i}}k},
  {and} \bibinfo{person}{Kemal Ebcioglu}.} \bibinfo{year}{2005}\natexlab{}.
\newblock \showarticletitle{Programming by sketching for bit-streaming
  programs}. \bibinfo{publisher}{ACM}, \bibinfo{pages}{281--294}.
\newblock


\bibitem[\protect\citeauthoryear{Solar-Lezama, Tancau, Bodik, Seshia, and
  Saraswat}{Solar-Lezama et~al\mbox{.}}{2006}]%
        {solar2006combinatorial}
\bibfield{author}{\bibinfo{person}{Armando Solar-Lezama},
  \bibinfo{person}{Liviu Tancau}, \bibinfo{person}{Rastislav Bodik},
  \bibinfo{person}{Sanjit Seshia}, {and} \bibinfo{person}{Vijay Saraswat}.}
  \bibinfo{year}{2006}\natexlab{}.
\newblock \showarticletitle{Combinatorial Sketching for Finite Programs}.
  \bibinfo{publisher}{ACM}, \bibinfo{pages}{404--415}.
\newblock


\bibitem[\protect\citeauthoryear{Wang, Cheung, and Bodik}{Wang
  et~al\mbox{.}}{2017a}]%
        {wang2017interactive}
\bibfield{author}{\bibinfo{person}{Chenglong Wang}, \bibinfo{person}{Alvin
  Cheung}, {and} \bibinfo{person}{Rastislav Bodik}.}
  \bibinfo{year}{2017}\natexlab{a}.
\newblock \showarticletitle{Interactive query synthesis from input-output
  examples}. In \bibinfo{booktitle}{\emph{Proceedings of the 2017 ACM
  International Conference on Management of Data}}. ACM,
  \bibinfo{pages}{1631--1634}.
\newblock


\bibitem[\protect\citeauthoryear{Wang, Cheung, and Bodik}{Wang
  et~al\mbox{.}}{2017b}]%
        {wang2017synthesizing}
\bibfield{author}{\bibinfo{person}{Chenglong Wang}, \bibinfo{person}{Alvin
  Cheung}, {and} \bibinfo{person}{Rastislav Bodik}.}
  \bibinfo{year}{2017}\natexlab{b}.
\newblock \showarticletitle{Synthesizing highly expressive SQL queries from
  input-output examples}. ACM, \bibinfo{pages}{452--466}.
\newblock


\bibitem[\protect\citeauthoryear{Yaghmazadeh, Wang, and Dillig}{Yaghmazadeh
  et~al\mbox{.}}{2018}]%
        {yaghmazadeh2018automated}
\bibfield{author}{\bibinfo{person}{Navid Yaghmazadeh}, \bibinfo{person}{Xinyu
  Wang}, {and} \bibinfo{person}{Isil Dillig}.} \bibinfo{year}{2018}\natexlab{}.
\newblock \showarticletitle{Automated migration of hierarchical data to
  relational tables using programming-by-example}.
\newblock \bibinfo{journal}{\emph{Proceedings of the VLDB Endowment}}
  \bibinfo{volume}{11}, \bibinfo{number}{5} (\bibinfo{year}{2018}),
  \bibinfo{pages}{580--593}.
\newblock


\bibitem[\protect\citeauthoryear{Yaghmazadeh, Wang, Dillig, and
  Dillig}{Yaghmazadeh et~al\mbox{.}}{2017}]%
        {yaghmazadeh2017sqlizer}
\bibfield{author}{\bibinfo{person}{Navid Yaghmazadeh}, \bibinfo{person}{Yuepeng
  Wang}, \bibinfo{person}{Isil Dillig}, {and} \bibinfo{person}{Thomas Dillig}.}
  \bibinfo{year}{2017}\natexlab{}.
\newblock \showarticletitle{{SQLizer: Query Synthesis from Natural Language}}.
  \bibinfo{publisher}{ACM}, \bibinfo{pages}{63:1--63:26}.
\newblock


\end{thebibliography}

\clearpage
\appendix
\section{Proofs}
\label{sec:appproofs}

\paragraph{Proof of Theorem~\ref{thm:main}.}

The key invariant maintained by our algorithm is that the true program is a completion of our current sketch $P$---i.e., $P\xRightarrow{*}\complete{P}^*$. We assume that the user provides a valid initial sketch, so this invariant holds at the beginning. Then, assuming the user answers queries correctly, if $\OO(\hat Q)=\true$, we know that $\complete{P}^*$ is a completion of $\hat Q$; thus, the update $P\gets\hat Q$ maintains this invariant. As a consequence, we guarantee that our algorithm returns $P=\complete{P}^*$. In particular, at this point, $P$ is both complete and satisfies $P\xRightarrow{*}\complete{P}^*$, but the only completion of a complete sketch is itself; thus, $P=\complete{P}^*$. $\qed$

\paragraph{Proof of Theorem~\ref{thm:complete}.}

First, note that the number of successful iterations (i.e., iterations where $\OO(\hat{Q})=\true$) is at most $|\complete{P}^*|$, since each successful iteration adds a node to the current sketch $P$, and we can add at most $|\complete{P}^*|$ nodes total.

Next, the number of unsuccessful iterations before a successful iteration is at most the number of queries $|\Q_P|$, where $P$ is the current sketch. To bound $|\Q_P|$, note that for each hole $N$ in $P$, if $A_N=I$, then there are at most $n$ ways to fill that hole (where $n$ is the number of tables in the database), and if $A_N=C$, then there are at most $m$ ways to fill that hole (where $m$ is the number of columns in the database). Finally, there are at most $|\complete{P}^*|$ holes in $P$. Thus, we have $|\Q_P|\le(n+m)\cdot|\complete{P}^*|$. As a consequence, the total number of iterations is $O((n+m)\cdot|\complete{P}^*|^2)$, as claimed. $\qed$

\section{Implementation Details}
\label{sec:appimpl}

We give details on our implementation.

\paragraph{Restriction to joins on keys.}

To constrain the search space, we restrict to inner-join operations on column pairs with matching keys (both primary key-foreign key joins and foreign key-foreign key joins). This restriction both improves performance and reduces the number of iterations needed.

\paragraph{Modified candidate queries.}

One challenge with our candidate queries $\Q_P$ is that if $P$ contains an expression $t_i\Join_{C,C}I$, then one of the constructed queries is
\begin{align*}
t_i\Join_{C,C}I\xRightarrow{*}t_i\Join_{C,C}t_j,
\end{align*}
for some table $t_j$. However, it might be hard for a user decide if this refinement is correct without knowing in advance the values of the column holes $C,C$ on which $t_i$ and $t_j$ are joined. Two other constructed queries are
\begin{align*}
t_i\Join_{C,C}I&\Rightarrow t_i\Join_{c,C}I \\
t_i\Join_{C,C}I&\Rightarrow t_i\Join_{C,c}I,
\end{align*}
where $c$ is a column; as before, it might be hard for the user to know if either of these refinements are correct without knowing the other table $I$ and the other column $C$. Thus, we require that these decisions be made together---i.e., for any expression of the form $t_i\Join_{C,C}I$, we only consider refinements of the form
\begin{align*}
t_i&\Join_{C,C}I\xRightarrow{*}t_i\Join_{c,c'}t_j \\
t_i&\Join_{C,C}I\xRightarrow{*}t_i\Join_{c,c'}t_j\Join_{C,C}I
\end{align*}
This modification changes Theorem~\ref{thm:complete}---in particular, the dependence of the maximum possible number of iterations on the number of columns $m$ and the number of tables $n$ changes. Nevertheless, the number of iterations remains polynomial in these parameters.

\paragraph{Type constraints.}

So far, we have largely ignored the fact that values $x$ in the sketch have types (i.e., integers, floats, strings, and regular expressions). Similarly, values in the database also have types (i.e., integers, floats, and strings). We can use these types to prune the space of programs. In particular, we impose these constraints on our grammar---i.e., in the expressions, $X\in C$ and $C~U~X$, we require that the type of $C$ and $X$ be identical. A special case is when $X$ is a regular expression, in which case we require that $C$ have type string; in this case, we also require that $U$ to be $\simeq$. These constraints implicitly affect both our algorithm when constructing queries and when sampling completions.

\paragraph{Candidate queries from samples.}

We only consider candidate queries $Q$ such that $Q\xRightarrow{*}\complete{P}$ for some sampled completion $\complete{P}\in\PP$ of the current sketch $P$. If $Q$ does not satisfy this property, then it has estimated score $\scorehat{Q;\PP}$. In particular, $\hat\pi_-=0$ since $\mathbb{I}[Q\xRightarrow{*}\complete{P}]=0$ for every $\complete{P}\in\PP$. In other words, no sampled completion would lead to the user responding $\OO(Q)=\true$. Thus, we can safely ignore it.

\paragraph{Precomputing soft constraints.}

Note that the rules for scoring completions $\complete{P}$ (i.e., computing $\semantics{\complete{P}}_{\varphi}$ in Figure~\ref{fig:soft} rely on the semantics $\semantics{\cdot}$ of subexpressions of $\complete{P}$. As part of our sampling procedure, we need to compute the score for a large number of completions $\complete{P}$. However, evaluating $\semantics{\complete{P}}$ can be computationally expensive.

Thus, we use an approximate approach to evaluating $\semantics{\cdot}_{\varphi}$. At a high level, for a given database and user-provided sketch $P$, we precompute the values of the soft constraints $\phi$ in $P$ on every column $c$ in the database. We assume tha columns are unique (i.e., no two tables have columns with the same name); we can achieve uniqueness by renaming columns. Then, instead of using $\semantics{\cdot}$, we use semantics $\semantics{\cdot}_{\approx}$ that only keeps track of the columns in tables. To apply a soft constraint $\phi$, we use the precomputed value if the corresponding column if it is in the table, and use $-\infty$ otherwise.

In more detail, we precompute values for every \emph{primitive soft constraint} $\phi$ in the original sketch $P$---i.e., $\phi=x\in c$, and $\phi=c~u~x$ for some $u\in\{\lesssim,\simeq,\gtrsim\}$. There are two cases. First, if $c$ is a column constant, then we precompute the value
\begin{align*}
\theta_{\phi}=\semantics{\phi}_{\varphi}~t,
\end{align*}
where $t$ is the table that contains $c$. Second, if $c$ in a hole, then for some column constant $c'$, we let $\phi_{c'}$ be the expression obtained by filling hole $c$ in $\phi$ with the production $c\Rightarrow c'$. Note that $\phi_{c'}$ cannot have any more holes, since $x$ and $u$ are not allowed to be holes. Then, we precompute
\begin{align*}
\theta_{\phi_{c'}}=\semantics{\phi_{c'}}_{\varphi}~t_{c'}
\end{align*}
for every column $c'$ in the database, where $t_{c'}$ is the table containing $c'$. As an example, in $P_{\text{author}}$, there are three primitive soft constraints $C:\texttt{c\_name}\simeq2$, $C:\texttt{c\_year}\gtrsim1900$, and $C:\texttt{c\_year}\lesssim2020$. For the first constraint $C:\texttt{c\_name}\simeq2$, we precompute $\theta_{\texttt{name}\simeq2}=0.5$, $\theta_{\texttt{title}\simeq2}=0.33$, etc. The others are similar. We also impose type constraints as described above for the modified candidate queries.

Then, we define the following approximate semantics for evaluating tables, which only keeps track of the columns in each table, and ignores the actual rows in the table:
\begin{align*}
\semantics{\Pi_{c_1,...,c_k}(s)}_{\approx}&=\semantics{s}_{\approx}\setminus\{\semantics{c_1},...,\semantics{c_r}\} \\
\semantics{\sigma_{\psi}(i)}_{\approx}&=\semantics{i}_{\approx} \\
\semantics{t}_{\approx}&=\alpha(t) \\
\semantics{t\Join_{c,c'}i}_{\approx}&=(\semantics{t}_{\approx}\setminus\{\semantics{c}\})\cup(\semantics{i}_{\approx}\setminus\{\semantics{c'}\})
\end{align*}
where $\alpha:t\mapsto(c_1,...,c_k)$ maps a table to its columns.

Finally, we correspondingly modify the soft constraint semantics $\semantics{\cdot}_{\varphi}$ to obtain an approximate version $\semantics{\cdot}_{\varphi,\approx}$. These semantics are identical to the semantics in Figure~\ref{fig:soft}, except (i) $\semantics{\cdot}$ is replaced with $\semantics{\cdot}_{\approx}$, and (ii) for the primitive soft constraints $x\in c$ and $x~u~c$, we use the rules
\begin{align*}
\semantics{x\in c}_{\varphi,\approx}&=\lambda t\;.\;\text{if }c\in t\text{ then }\theta_{x\in c}\text{ else }-\infty \\
\semantics{x~u~c}_{\varphi,\approx}&=\lambda t\;.\;\text{if }c\in t\text{ then }\theta_{x\in c}\text{ else }-\infty
\end{align*}
in place of the ones in Figure~\ref{fig:soft}.

This approximation actually changes the semantics of soft constraints. For example, for $\complete{P}_{\text{author}}$, we have
\begin{align*}
\semantics{\complete{t}}_{\approx}=(\text{aid},\text{name},\text{pid},\text{title},\text{year}),
\end{align*}
where $\complete{t}$ is the subexpression of $\complete{P}_{\text{author}}$. Then, for the primitive soft constraint $\texttt{name}\simeq2$ in the soft constraints $\complete{\phi}$ in $\complete{P}_{\text{author}}$, we have
\begin{align*}
\semantics{\texttt{name}\simeq2~\complete{t}}_{\varphi,\approx}=\theta_{\texttt{name}\simeq2}=0.5,
\end{align*}
since $\texttt{name}\in\semantics{\complete{t}}_{\approx}$. However, recall that $\semantics{\texttt{name}\simeq2~\complete{t}}=0.33$, which shows that the semantics are different.

Intuitively, the difference is that $\semantics{\cdot}_{\varphi}$ is evaluated on the column observed during execution, whereas $\semantics{\cdot}_{\varphi,\approx}$ is evaluated on the original column. During execution, values in the column can be duplicated or deleted---e.g., due to inner-join operations with other columns or select operations. For example, in Figure~\ref{fig:exdb}, the value ``Alan M. Turing'' is duplicated since he has two papers. Indeed, in some cases, $\semantics{\cdot}_{\varphi,\approx}$ may actually be more intuitive compared to $\semantics{\cdot}_{\varphi}$.

\paragraph{Modified scoring function.}

When scoring programs $\complete{P}$, we also add a term based on the size of $\complete{P}$; we measure the size of $\complete{P}$ in terms of number of nodes in its AST, which we denote $|\complete{P}|$. In particular, we use
\begin{align*}
\scoretilde{Q;\pi_P}=\score{Q;\pi_P}+\lambda\cdot|\complete{P}|,
\end{align*}
where $\lambda\in\mathbb{R}_{\ge0}$ is a hyperparameter. We make a similar modification to $\scorehat{Q;\pi_P}$. In addition, we assign a score of $-\infty$ to $\complete{P}$ if there is some table operation in $\complete{P}$ for which a column in that operation is not contained in the corresponding table. In particular, (i) for a project operation $\Pi_{c_1,...,c_k}(t)$, we must have $c_1,...,c_k\in\semantics{t}_{\approx}$, (ii) for a select operation $\sigma_{\psi}(t)$, any column $c$ appearing in $\psi$ must satisfy $c\in\semantics{t}_{\approx}$, and (iii) for an inner-join operation $t\Join_{c,c'}t'$, we must have $c,c'\in\semantics{t}_{\approx}$.

\end{document}